\documentclass[prb,twocolumn,amsmath,amssymb,floatfix]{revtex4-2}
\usepackage{epsfig, graphicx,graphics,amsmath,amssymb,float}
\usepackage[T1]{fontenc}
\usepackage[latin9]{inputenc}
\usepackage{amsmath}
\usepackage{amssymb}
\usepackage{appendix}
\usepackage{amscd}
\usepackage{bm}
\usepackage{psfrag}
\usepackage{bbm} 
\usepackage{babel}
\usepackage{wasysym}
\usepackage{mathrsfs}
\usepackage{color}
\usepackage[normalem]{ulem}
\usepackage{tabularx} 
\usepackage{tikz}
\usetikzlibrary{calc,decorations.markings}
\usepackage[final]{hyperref} 
\hypersetup{
	colorlinks=true,       
	linkcolor=blue,        
	citecolor=blue,        
	filecolor=magenta,     
	urlcolor=blue         
}

\newcommand{\new}[1]{\textcolor{black}{#1}}

\usepackage{mathtools}
\DeclarePairedDelimiter\bra{\langle}{\rvert}
\DeclarePairedDelimiter\ket{\lvert}{\rangle}
\DeclarePairedDelimiterX\braket[2]{\langle}{\rangle}{#1 \delimsize\vert #2}

\begin{document}

\title{Many-body quantum dynamics of spin-orbit coupled Andreev states in a Zeeman field}

\author{Kateryna Zatsarynna}
\affiliation{Institut f\"ur Theoretische Physik, Heinrich-Heine-Universit\"at, D-40225  D\"usseldorf, Germany}
\author{Andrea Nava}
\affiliation{Institut f\"ur Theoretische Physik, Heinrich-Heine-Universit\"at, D-40225  D\"usseldorf, Germany}
\author{Alex Zazunov}
\affiliation{Institut f\"ur Theoretische Physik, Heinrich-Heine-Universit\"at, D-40225  D\"usseldorf, Germany}
\author{Reinhold Egger}
\affiliation{Institut f\"ur Theoretische Physik, Heinrich-Heine-Universit\"at, D-40225  D\"usseldorf, Germany}

\begin{abstract}
We provide a theoretical framework to describe the quantum many-body dynamics of Andreev states in
Josephson junctions with spin-orbit coupling and a magnetic Zeeman field.  In such cases, employing a doubled Nambu spinor description is 
technically advantageous but one then has to be careful to avoid double-counting problems.  
By deriving the Lindblad master equation in the so-called excitation picture, we show that a physically consistent many-body theory free from double-counting problems follows.  We apply our formalism to a study of dynamical parity stabilization of the Andreev sector at intermediate times after an initial microwave pulse, in particular addressing the combined effects of spin-orbit coupling and Zeeman field. 
\end{abstract}
\maketitle

\section{Introduction} \label{sec1}

At present, nanoscale Josephson junctions are intensely studied in view of their relevance for many applications, e.g., in 
quantum information processing, as ultra-sensitive quantum sensors, or as superconducting diodes \cite{Alvaro2011,Wendin2017,Kjaergaard2020,Blais2021,Rasmussen2021,Nadeem2023}.
Over the past decade, high-quality hybrid nanowires realizing Josephson junctions with just a few transport channels of high transmission probability have become available in different laboratories \cite{Zgirski2011,Bretheau2013,Janvier2015,Larsen2015,Lange2015,Woerkom2017,Hays2018,Tosi2019,Hays2020,Whiticar2021,Hays2021,Metzger2021,Fatemi2022,Poschl2022,Danilenko2023,Wesdorp2023,Bargerbos2023,Pita2023,Wesdorp2024,Driel2024}.  Typically, the supercurrent is then mostly  carried by subgap Andreev bound states (ABSs) \cite{Kalenkov2009,Alvaro2011,Beenakker1991,Furusaki1991,Bagwell1992,Desposito2001,Nazarov2009,Kurilovich2021} 
localized near the weak link between the superconducting banks.  
Importantly, these ABSs can also be used to encode a qubit degree of freedom \cite{Zazunov2003,Nazarov2003,Zazunov2005,Padurariu2010,Padu2012} if one can preserve the fermion number parity  of the Andreev sector (simply referred to as ``parity'' below) on sufficiently long time scales below the parity switching time $\tau_p$. The  time scale $\tau_p$ describing transitions between states of 
opposite parity is generated by a variety of microscopic mechanisms \cite{Wesdorp2023,Ackermann2023,Kurilovich2024,Sahu2024}.  Recent experiments have shown that coherent Andreev qubit manipulations are feasible on time scales of up to
$\sim 100\mu$s \cite{Janvier2015,Hays2018,Hays2020,Hays2021,Wesdorp2023,Bargerbos2023,Pita2023,Wesdorp2024}.  

In the context of quantum information processing applications, spin-based Andreev qubits \cite{Brunetti2013,vanHeck2017,Padurariu2010,Bargerbos2023,Pita2023,Wesdorp2023,Wesdorp2024} are of particular importance. Here spin-orbit interaction (SOI) effects in combination with weak magnetic Zeeman fields play a central role.  The corresponding qubit manipulations are possible through electrostatic gate modulations of the SOI \cite{Tosi2019,Metzger2021,Matute2022,Pita2023},
by magnetic flux variations \cite{Hays2020,Hays2021}, and/or by Zeeman field changes.
In fact,  many nanowires studied experimentally so far are based on material platforms with strong SOI, e.g., InAs or InSb. 
We here study weak links of intermediate length $L\approx\xi_0$, where $\xi_0$ is the superconducting coherence length.  One then finds typically four (spin-split) positive-energy ABSs and nontrivial consequences of the SOI can arise. 
(In the short-junction limit $L\ll \xi_0$ \cite{Kos2013,Zazunov2014,Olivares2014,Riwar2015,Park2020}, there are only two levels and SOI does not cause new physics. For the complementary long-junction limit, see, e.g., Refs.~\cite{Giuliano_2013,Giuliano_2014,Nava_2016}.)
Below we mainly consider Josephson junctions with relatively high transparency, where electron-electron interaction effects are strongly suppressed; we therefore neglect interaction effects, but see Refs.~\cite{Matute2022,Buccheri2022,Giuliano2022a}.

Previous theory work has analyzed the ABS dispersion relation and the corresponding wave functions in Josephson junctions with SOI and Zeeman fields \cite{Park2017,vanHeck2017,Campagnano_2015,Minutillo2018,Matute2022,Fauvel2024,Lidal2024}. Here we go beyond those works and study the \emph{nonequilibrium population dynamics} in the Andreev sector in the presence of both SOI and Zeeman field. The analogous case without SOI and magnetic field has been studied in Ref.~\cite{Ackermann2023}.  We derive the master equation governing the 
population dynamics for the present case. We then apply the formalism to investigate the impact of SOI and magnetic field on the Andreev
population dynamics after an initial microwave pulse.  Such a pulse has been shown experimentally \cite{Wesdorp2023} and theoretically \cite{Ackermann2023,Kurilovich2024} to allow for dynamical
parity polarization over long but finite time scales.   We here examine how this phenomenon is affected
by the SOI and the Zeeman field.  

The structure of the remainder of this paper is as follows.  In Sec.~\ref{sec2}, we describe our model.  For details on the eigenstates, we refer to the Appendix.  In Sec.~\ref{sec3}, we derive a Lindblad master equation governing the dynamics of the Andreev sector (under certain assumptions specified below).  
The diagonal elements of the time-dependent reduced density matrix describing the Andreev sector, which are associated with the population probabilities of many-body Andreev states, obey a matrix rate equation which we specify explicitly.
In Sec.~\ref{sec4}, we then use this matrix rate equation to study the population dynamics
after an initial strong microwave pulse.  We compare our results to those of Ref.~\cite{Ackermann2023}, obtained in the absence of the SOI and the Zeeman field.  Importantly, we do not attempt a quantitative comparison to the
experiments of Ref.~\cite{Wesdorp2023}.  Instead our main goal is to provide a conceptual framework for describing
the many-body population dynamics in the Andreev sector if both the SOI and a Zeeman field are present.  In such cases,
it is technically convenient to work in an augmented space where Nambu spinor fields are doubled \cite{Hasan2010}.
However, one must then make sure that no double-counting problems arise. We here show how to consistently 
formulate the Lindblad equation approach in such cases. Our formalism is generally applicable for this type of problem, well
beyond the specific example discussed in this work.   Other applications of this approach will be reported elsewhere.
Finally, we conclude with a summary and an outlook in Sec.~\ref{sec5}.

\section{Model}\label{sec2}

\begin{figure}
\begin{center}
    \includegraphics[width=0.4\textwidth]{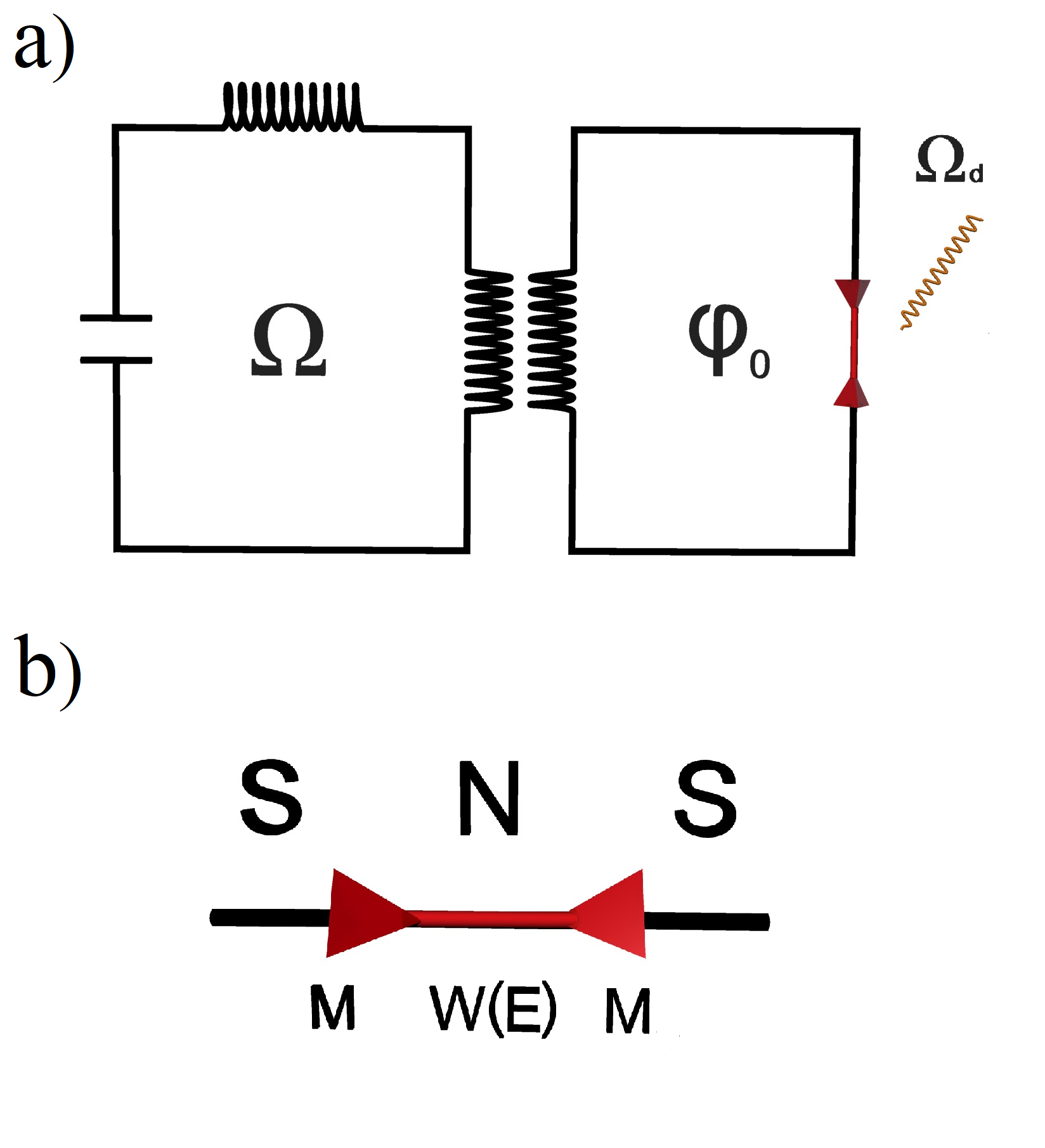}
    \caption{Schematic sketch of the studied setup. (a) A Josephson junction is embedded in a superconducting loop threaded by a  magnetic flux and inductively coupled to a microwave resonator with resonance frequency $\Omega$.  The junction can be driven by a microwave pulse of frequency $\Omega_d$.  The average phase difference across the junction is $\varphi_0$, and we assume a homogeneous pairing gap $\Delta$ in the loop containing the junction.
    (b) The Josephson junction is formed by a ballistic spinful single-channel nanowire of length $L$ between two superconducting banks. At the left and right ends of the wire, boundary states in the respective superconductor and in the wire are coupled by spin- and energy-independent tunneling amplitudes which we assume to be equal. These are encoded by the matrix $M$. 
    We also include spin-orbit coupling and a constant magnetic Zeeman field in the nanowire, encoded by the matrix $W(E)$. For details, see main text. }
    \label{fig1}
\end{center}
\end{figure}

In this section, we describe the model used in our study, see Sec.~\ref{sec2a}, where a spin-orbit coupled Josephson 
junction is embedded in a superconducting loop coupled to an electromagnetic environment.  In Sec.~\ref{sec2b}, we perform
an expansion to lowest order in the coupling to the environment to arrive at the model studied in the remainder of the paper.
Unless specified explicitly, we use units with $\hbar=e=k_B=1$ throughout the paper.  The notation $0^\pm$ implies positive and negative infinitesimals, respectively.

\subsection{Josephson junction with spin-orbit coupling and Zeeman field}\label{sec2a}

We consider a single-channel Josephson junction between two superconductors of conventional
$s$-wave BCS type with the same pairing gap $\Delta$ and the same Fermi velocity $v_F$. The coherence length is then given by $\xi_0=v_F/\Delta$. The respective order parameter phases are denoted by $\phi_1$ and $\phi_2$. The weak link representing the junction region is assumed to be a normal-conducting impurity-free one-dimensional (1D) nanowire of length $L$. This single-channel wire is connected by 
tunnel couplings at its ends ($x=\mp L/2$) to the respective superconducting bank, see Fig.~\ref{fig1}(b).
In the nanowire region, we include the SOI, where the polar axis defines the $z$-direction and the SOI strength is encoded by a parameter $\gamma^{}_{\rm SO}$. We also include a weak magnetic Zeeman field $\propto {\bf b}$, see Eq.~\eqref{Hn} below.  (In the superconducting banks, 
this field may slightly renormalize $\Delta$. This effect is kept implicit below.)  
For concreteness, as in Ref.~\cite{Ackermann2023}, we assume that the Josephson junction is embedded in a loop inductively coupled to a microwave resonator with resonance frequency $\Omega$, see Fig.~\ref{fig1}(a).  This resonator is responsible for an electromagnetic 
environment that triggers transitions between the fermionic eigenstates of the junction. (Our formalism can easily be adapted to other types
of electromagnetic environments.) In addition, a magnetic
flux threading the superconducting loop containing the weak link imposes the average phase difference $\varphi_0=\phi_1-\phi_2$ across the Josephson 
junction.  

Within the standard low-energy quasi-classical theory approach \cite{Nazarov2009}, one describes the superconductors in terms of 
field envelopes $\Psi_{\alpha,\sigma}(x,t)$ for right- or left-moving ($\alpha=\pm$) electrons with 
spin $\sigma\in \{\uparrow,\downarrow\}$. With the coordinate $x<0$ ($x>0$) for the left (right) superconductor, we here retain 
only the 1D channel propagating through the junction and perform a low-energy expansion around the Fermi momenta $\pm k_F$. 
The above fields are collected into a single four-spinor field,
\begin{equation}\label{Psidef}
    \Psi(x,t) = \left( \begin{array}{c}   \Psi_{+, \uparrow}\\    \Psi_{+, \downarrow}\\
    \Psi_{-, \uparrow} \\ \Psi_{-, \downarrow}    \end{array}\right).
\end{equation}
To efficiently account for the SOI and the Zeeman field in the normal region, we define an
eight-spinor field $\Phi(x,t)$ by employing particle-hole (Nambu) space \cite{Hasan2010}, 
\begin{equation}\label{Nambu}
    \Phi(x,t) = \frac{1}{\sqrt{2}} \begin{pmatrix} \Psi(x,t) \\  \tilde{\Psi}^*(x,t)
    \end{pmatrix},
\end{equation}
with
\begin{equation}
    \tilde{\Psi}(x,t) = \rho_x i\sigma_y \Psi(x,t) = 
     \left( \begin{array}{c}   \Psi_{-, \downarrow}\\    -\Psi_{-, \uparrow}\\
    \Psi_{+, \downarrow} \\ -\Psi_{+, \uparrow}    \end{array}\right).
\end{equation}
We use Pauli matrices $\tau_{x,y,z}$ in Nambu space, $\rho_{x,y,z}$ in 
right-left mover space, and $\sigma_{x,y,z}$ in spin space. The corresponding identity matrices
$(\tau_0,\rho_0,\sigma_0)$ are often kept implicit.
The Nambu spinor field in Eq.~\eqref{Nambu} satisfies the reality constraint 
\begin{equation}\label{reality}
    \rho_x \sigma_y \tau_y \Phi(x,t) = \Phi^*(x,t),
\end{equation} 
which implies redundancy. Hence one needs to be careful to avoid double counting problems \cite{Hasan2010}.

With the above definitions, within the low-energy quasi-classical approximation, the superconducting banks are described by a Bogoliubov-de Gennes (BdG) 
Hamiltonian \cite{Nazarov2009,Ackermann2023}, 
\begin{eqnarray} \nonumber
    H(t) &=& \sum_{j=1,2}\int_{s_j x<0} dx\, \Phi^\dagger(x,t) \Bigl[ \left(-iv_F \rho_z \partial_x 
    + V_j(t) \right)\tau_z\\ \label{BCS} && \qquad\quad  +\, \Delta \tau_x e^{i\tau_z \phi_j(t)} \Bigr] \Phi(x, t),
\end{eqnarray}
where $V_j(t) = \dot{\phi}_j(t)/2$ follows from the second Josephson relation and we define $s_1=+1$ and $s_2=-1$.  
Note that the left superconductor ($j=1$) corresponds to $x<0$ and the right one ($j=2$) to $x>0$.
The boundary Nambu spinor states $\Phi(0^-,t)$ and $\Phi(0^+,t)$ are then
tunnel-coupled to the respective ends of the normal wire forming the weak link. We next show that those couplings generate a time- (or energy-)dependent transfer matrix connecting these boundary spinors.

The single-particle Hamiltonian for the uncoupled and ballistic (impurity-free) normal-conducting nanowire of length $L$, with new coordinates $|x|<L/2$ pertaining to the nanowire, is taken in the form
\begin{equation} \label{Hn}
    h_N = \frac{\hat p^2}{2m} + \gamma_{\rm SO}^{} \,\hat p\, \sigma_z + {\bf b} \cdot {\bm \sigma},
\end{equation}
with an effective mass $m$ and the 1D momentum operator $\hat p$.  The SOI strength and the Zeeman field are 
encoded by $\gamma_{\rm SO}^{}$ and the vector ${\bf b}$, respectively, where 
${\bm \sigma}=(\sigma_x,\sigma_y,\sigma_z)$.  Estimates for $\gamma_{\rm SO}^{}$ for realistic geometries
can be found, e.g., in Refs.~\cite{Tosi2019,Lidal2024}.
We now linearize Eq.~\eqref{Hn} around the Fermi points in the wire, which are denoted by $\pm k_0$.  Using
$\hat p \rightarrow \alpha k_0 -i\partial_x$, we introduce field operators
$\psi^{}_{\alpha,\sigma}(x,E)$ for right- and left-movers ($\alpha=\pm$) in the nanowire
with spin $\sigma\in\{\uparrow,\downarrow\}$ and energy $E$.  With $\psi_\alpha=(\psi_{\alpha,\uparrow},\psi_{\alpha,\downarrow})^T$,
the second-quantized low-energy Hamiltonian for the nanowire is then given by
\begin{eqnarray}\nonumber
    H_N &=&  \sum_{\alpha=\pm} \int_{-L/2}^{L/2}dx\,\psi^\dagger_\alpha(x) \Bigl
    \{ [\alpha v_0 + \gamma_{\rm SO}^{}\sigma_z] \,(-i\partial_x) \\ \label{HNN}
    && +\, \alpha \gamma_{\rm SO}^{}\, k_0 \sigma_z + {\bf b} \cdot {\bm \sigma}\Bigr\} \psi_\alpha^{}(x),
\end{eqnarray}
where $v_0=k_0/m$ is the Fermi velocity in the wire. 
Since there is no backscattering inside the nanowire described by Eq.~\eqref{HNN}, 
the right-left-mover index $\alpha=\pm$ is conserved. 
We can thus connect the boundary spinors $\psi_\alpha(-L/2,E)$ and 
$\psi_\alpha(+L/2,E)$ by a transfer matrix $W_\alpha(E)$
in spin space.  Explicitly, we find
\begin{eqnarray} \label{transf1}
    \psi_{\alpha}(-L/2,E) &=& W_\alpha(E)\, \psi_{\alpha}(L/2,E),\\
    W_\alpha(E) &=& B_{\alpha}^{}
    \begin{pmatrix}
        e^{i\lambda_{\alpha,  \uparrow} L} & 0 \\
        0 & e^{i\lambda_{\alpha, \downarrow} L} 
    \end{pmatrix}   B^{-1}_{\alpha}, \nonumber
\end{eqnarray}
where the matrix $B_\alpha(E)$ and the numbers $\lambda_{\alpha,\sigma}(E)$ are found by diagonalizing a matrix resulting from Eq.~\eqref{HNN},
\begin{equation}
 \begin{pmatrix}
        \frac{b_z + \alpha \gamma_{\rm SO}^{} k_0-E}{\alpha v_0 + \gamma_{\rm SO}^{}} & \frac{b_x - ib_y}{\alpha v_0 + \gamma_{\rm SO}^{}} \\[1ex]
        \frac{b_x + ib_y}{\alpha v_0 - \gamma_{\rm SO}^{}} & 
        \frac{b_z + \alpha \gamma_{SO}^{} k_0+E}{\alpha v_0 - \gamma_{\rm SO}^{}} 
    \end{pmatrix}=  B_{\alpha}
    \begin{pmatrix} \lambda_{\alpha, \uparrow} & 0 \\   0 & \lambda_{\alpha,  \downarrow}
    \end{pmatrix} B^{-1}_{\alpha}.
\end{equation}
Below we use the four-spinor field $\psi(x) =(\psi_+,\psi_-)^T$, in analogy to the corresponding 
definition in the superconducting banks, see Eq.~\eqref{Psidef}. Furthermore, we use
a $4\times 4$ transfer matrix $W(E)$ which is diagonal in 
left-right-mover space, $W(E)={\rm diag}[W_+(E),W_-(E)]$.

Next we take into account spin-independent tunneling amplitudes connecting
the nanowire ends to the corresponding left and right superconducting banks. 
We assume that the respective contacts have the energy-independent
transmission probabilities ${\cal T}_1$ and ${\cal T}_2$.  For simplicity, in what
follows, we assume equal transmission probabilities, ${\cal T}_1={\cal T}_2={\cal T}$. However, the generalization 
to asymmetric cases poses no conceptual challenge. 
The corresponding reflection amplitude at each junction is then defined by $r=\sqrt{1-{\cal T}}$.  
At the left contact ($j=1$), the state at the right boundary of the left superconductor, $\Psi(0^-,E)$, and the state at
the left end of the nanowire, $\psi(-L/2,E)$, are then matched according to 
the transfer matrix condition \cite{Nazarov2009,Zazunov2005,Zazunov2014}
\begin{equation}\label{match1}
 \Psi(0^-,E) = M \psi(-L/2, E),\quad
 M = \frac{1}{\sqrt{\mathcal{T}}} (\rho_0 +r\rho_x) \sigma_0.
\end{equation}
We emphasize again our convention that the left (right) superconductor has spatial coordinates with $x<0$ ($x>0$), while we use different coordinates with $-L/2<x<L/2$ for the nanowire. Similarly, at the right contact ($j=2$), we have the condition
\begin{equation}\label{match2}
 \psi(L/2,E) =  M \Psi(0^+, E).
\end{equation}
Combining Eqs.~\eqref{match1} and \eqref{match2} with the transfer matrix $W(E)$
across the normal-conducting nanowire region,  we arrive at a 
matching condition connecting the two superconducting boundary states,
\begin{equation}\label{transfermatrix}
    \Psi(0^-, E) =  T(E)  \Psi(0^+, E), \quad
    T(E) = M W(E) M,
\end{equation}
where $T(E)$ is the full transfer matrix, see Fig.~\ref{fig1}(b). In this way, we have effectively integrated out the 
normal-conducting region.

The corresponding matching condition for the Nambu spinor states \eqref{Nambu} is given by
\begin{eqnarray}
 \label{matching}
    \Phi(0^-, E) &=& \hat{T}(E)\, \Phi(0^+, E), \\ \nonumber
    \hat{T}(E) &=& \begin{pmatrix}
        T(E) & 0 \\
        0 & \rho_x \sigma_y T^*(-E) \sigma_y \rho_x
    \end{pmatrix},    
\end{eqnarray}
where the explicit $2\times 2$ structure of $\hat{T}(E)$ refers to Nambu space.
In the time domain, the matching condition \eqref{matching} is equivalently written as 
\begin{equation}
    \Phi(0^-, t) = \hat{T}(t) \, \Phi( 0^+, t).
\end{equation}
Since we focus on the symmetric case ${\cal T}_1={\cal T}_2$, we are free to choose a gauge where the 
voltage in the normal-conducting region vanishes  and the superconducting phases 
can be written as $\phi_j(t)=s_j\varphi(t)/2$, with the phase difference 
$\varphi(t)$. One then obtains $\hat{T}(t)$ from $\hat{T}(E)$ through the  replacement $E\to i\partial_t$.
Finally, the Nambu spinors obey the normalization condition 
\begin{eqnarray} \label{normalization}
  &&  \int_{-\infty}^{\infty} dx\, \left|\Phi(x, E)\right|^2 = 1-\xi_w(E),\\
  \nonumber
  && \xi_w(E)=\frac{L}{2}\left(|\Phi(0^-,E)|^2+|\Phi(0^+,E)|^2\right).
\end{eqnarray}
The $\xi_w(E)$ term here arises due to the wave function weight in the normal-conducting region \cite{Ackermann2023}.

\subsection{Expansion in the system-environment coupling}\label{sec2b}

To proceed, we write the phase difference  as $\varphi(t) = \varphi_{0} + \delta\varphi(t)$, where the fluctuating phase $\delta \varphi(t)$ due to the microwave resonator is assumed to be 
a small perturbation, $|\delta\varphi(t)|\ll 1$.  Following Ref.~\cite{Ackermann2023}, 
we expand the BdG Hamiltonian to leading order in $\delta \varphi$. After a global canonical 
transformation,
\begin{equation}\label{globalgauge}
\left . \Phi(x,t) \right|_{s_j x<0} \, \rightarrow \,
e^{-i\tau_z s_j \varphi_0/4}\, \Phi(x,t) ,
\end{equation}
we obtain the Hamiltonian 
\begin{equation}\label{fullH}
H(t) = H_{0} + H_{I}(t) + H_{\rm env}+{\cal O}(\delta\varphi^2) ,
\end{equation}
where $H_{\rm env}$ describes the electromagnetic environment which is equivalent to a 
set of harmonic oscillators \cite{Nazarov2009}.
The noninteracting ($\delta\varphi=0$) BdG Hamiltonian is given by 
\begin{equation}
    H_{0} = \sum_{j=1,2} \int_{s_j x < 0} dx\, \Phi^\dagger(x,t) \left[-i v_F\rho_z\tau_z \partial_x  + 
    \Delta \tau_x \right] \Phi(x, t), \label{Hbdg}
\end{equation}
and the leading-order interaction term follows as
\begin{eqnarray} \nonumber
    H_{I}(t) &=& \sum_j \int_{s_j x < 0} dx\, \Phi^\dagger(x,t)\, \text{sgn}(-x)
    \Biggl[\frac{\delta \dot{\varphi}}{4} \tau_z + \\ \label{Hi} 
    && + \, \Delta \frac{\delta \varphi}{2} \tau_y \Biggr] \, \Phi(x, t).
\end{eqnarray}
Due to the transformation \eqref{globalgauge}, the transfer matrix 
acquires an additional phase factor.
As a result, the final matching condition reads
\begin{equation}\label{matchingfinal}
    \Phi(0^-, t) = e^{i\tau_z \varphi_0/2} \, \hat{T}(E\to i\partial_t) \, \Phi(0^+, t).
\end{equation}

We consider the above problem in the interaction picture. The Nambu 
field operator can be expanded in terms of the stationary eigenstates $\Phi_\nu(x)$ with energy $E_\nu$ 
of the BdG problem posed by
$H_0$ in Eq.~\eqref{Hbdg} and the matching condition \eqref{matchingfinal},
\begin{equation}\label{Phidef}
    \Phi(x,t) = \sum_{\nu} \Phi_{\nu}(x)\gamma_{\nu}(t),
\end{equation}
with fermion operators $\gamma_{\nu}(t) = e^{-iE_{\nu}t}\gamma_{\nu}$.  
Explicitly, with Eq.~\eqref{matching}, this BdG problem is given by  
\begin{eqnarray} \nonumber
    &&\left[ -iv_F\rho_z \tau_z\partial_x + \Delta \tau_x \right] \Phi_{\nu}(x) = 
    E_{\nu}\Phi_{\nu}(x), \\ \label{bdg}
    && \Phi_{\nu}(0^-) = e^{i\tau_z \varphi_0/2} \, \hat{T}(E_\nu) \, \Phi_{\nu}(0^+).
\end{eqnarray}
As a result,  we find
\begin{equation}\label{bdgf}
    H_0=\sum_\nu E_\nu \gamma_\nu^\dagger \gamma_\nu^{},
\end{equation}
where the index $\nu$ includes subgap ABS solutions with $|E_\nu|<\Delta$ as well as quasiparticle continuum states with quantum numbers $\nu=p\equiv (E,s,\sigma)$ where $|E|>\Delta$. The index $s\in\{1,2,3,4\}$ specifies the incoming scattering state type, and $\sigma$ refers to the spin state.  
Due to the particle-hole symmetry of the BdG Hamiltonian,  
\begin{equation}\label{phs}
{\cal C} H_0 {\cal C}^{-1} =- H_0,\quad {\cal C}=\sigma_y\tau_y K,
\end{equation}
where $K$ denotes complex conjugation, for every solution with 
$E_\nu>0$, we must have a corresponding solution at the opposite 
energy $E_{\bar\nu}=-E_\nu$.  This fact is readily shown by combining
Eqs.~\eqref{reality} and \eqref{bdg}, see also Ref.~\cite{Park2017}.  Using in addition Eq.~\eqref{Phidef}, 
one finds that the corresponding quasiparticle operator is given by $\gamma_{\bar \nu}^{}=\gamma_\nu^\dagger$.

In the interaction picture, up to an irrelevant time-derivative term \cite{Ackermann2023}, the interaction term \eqref{Hi} can be written as
\begin{equation}\label{HI}
   H_I(t) =  \frac{\delta \varphi(t)}{2} \mathcal{I}(t),
\end{equation}
with the Josephson current operator 
\begin{eqnarray}\nonumber
    \mathcal{I}(t) &=& \sum_{\mu\ne \nu} \mathcal{I}_{\mu ,\nu} \gamma^{\dagger}_{\mu}(t) \gamma_{\nu}^{}(t), \\
    \mathcal{I}_{\mu,\nu} &=&   \int dx\, \Phi^\dagger_{\mu}(x)\, {\rm sgn}(-x) \nonumber 
    \left[\frac{E_{\mu}-E_{\nu}}{2i} \tau_z  + \Delta \tau_y \right] \Phi_{\nu}(x)\\ &=& \mathcal{I}_{\nu,\mu}^{\,\ast} .\label{JC} 
\end{eqnarray}
We provide a concise discussion of the BdG eigenstates resulting from Eq.~\eqref{bdg} and of the 
current matrix elements \eqref{JC} in the Appendix.

In what follows, we denote ABS solutions with the quantum number $\nu=\lambda$.
Using the matrix $T(E)$ in Eq.~\eqref{transfermatrix} and the function $\gamma(E)=\cos^{-1}(E/\Delta)$, 
we show in the Appendix that the matching equation has nontrivial solutions only for energies satisfying
the condition
\begin{equation}\label{detzero}
   \text{det}[A_p(E)-A_h(E)]=0
\end{equation}
with the particle and hole matrices
\begin{eqnarray}\nonumber
    A_p(E) &=&  e^{i \varphi_0/2} e^{i \rho_z \gamma(E)/2} \, T(E) \, e^{i \rho_z \gamma(E)/2}, \\ 
    A_h(E) &=& \rho_x \sigma_y A_p^*(-E)\sigma_y \rho_x.
\end{eqnarray}
We obtain $E_\lambda(\varphi_0)$ and the corresponding ABS wave function, see Eq.~\eqref{ABSwf} in the Appendix,
by numerically solving Eq.~\eqref{detzero} and determining the corresponding eigenvectors.  
In practice, we study cases with $L\approx \xi_0=v_F/\Delta$, where one typically encounters four spin-split positive-energy ABS solutions.

\section{Many-body Andreev-state population dynamics}\label{sec3}

In this section, we derive the dynamical equations governing the time evolution of the many-body Andreev states
for the above model.  
First, in Sec.~\ref{sec3a}, we introduce the so-called excitation picture and contrast it with the 
alternative semiconductor picture \cite{Zazunov2014,Ackermann2023}.
We show that the excitation picture offers a particularly convenient representation for 
superconducting problems with SOI and Zeeman fields, since double-counting issues are more difficult to 
handle in the semiconductor picture. 
In Sec.~\ref{sec3b}, we then derive a Lindblad master equation for the dynamics of
the reduced density operator $\rho_A(t)$ describing the Andreev-state sector. 
In Sec.~\ref{sec3c}, we discuss the many-body population dynamics in the Andreev subspace by considering the diagonal elements of $\rho_A(t)$. For $\gamma_{\rm SO}^{}=0$ and ${\bf b}=0$, our approach recovers 
the results of Ref.~\cite{Ackermann2023}.  Applications of the formalism to cases with finite SOI and/or magnetic Zeeman field are presented in Sec.~\ref{sec4}.
 
\subsection{Excitation picture vs semiconductor picture}\label{sec3a}

The Nambu representation introduced in Eq.~\eqref{Nambu} is very convenient for 
theoretically handling the combined effects of superconductivity, SOI,  and Zeeman fields in a unified framework \cite{Hasan2010}. 
However, due to the reality constraint \eqref{reality}, this representation also comes at a cost since it implies an artificial doubling of the number of single-particle states. We explain below how one can circumvent the appearance of spurious non-physical many-body states in such a formulation.

\begin{figure}
\begin{center}
    \includegraphics[width=0.5\textwidth]{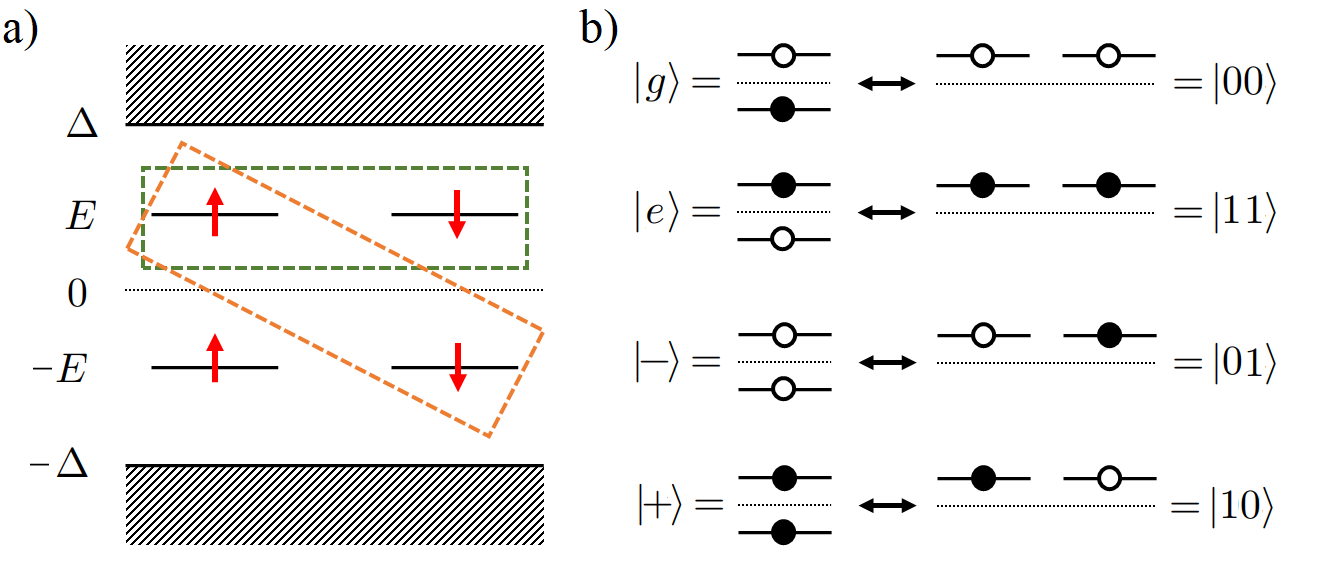}
    \caption{Schematic BdG spectrum of a short ($L\ll\xi_0)$ junction without SOI and Zeeman field, taken at fixed phase difference $\varphi_0$.
    In this case, one obtains a single spin-degenerate ABS with energies $E_\uparrow=E_\downarrow=E>0$ and $E<\Delta$, plus the corresponding states at energy $-E$ obtained from 
    particle-hole symmetry. (a) Double counting can be removed in the semiconductor picture by retaining only the 
    single-particle states $(E_\uparrow,-E_\downarrow)$ inside the tilted orange-dashed box.  
    Alternatively, in the excitation picture, double counting is removed by retaining only 
    the positive-energy single-particle states $(E_\uparrow,E_\downarrow)$ inside the horizontal green-dashed box. Continuum states with $|E|>\Delta$ correspond to the shaded regions.
    (b) The four possible many-body Andreev states $\{ |g\rangle,|e\rangle,|-\rangle,|+\rangle\}$ in the semiconductor picture (left side), where filled (empty) dots indicate occupied (empty) single-particle ABS levels.
    In the equivalent excitation picture (right side), these four states are represented by $\{ |00\rangle, |11\rangle, |01\rangle, |10\rangle\}$, respectively.}
    \label{fig2}
\end{center}
\end{figure}

Let us first consider the case of a short junction without SOI and Zeeman field at fixed phase difference $\varphi_0$ 
and fixed other parameters, see Fig.~\ref{fig2}.  In this example,  we have a single spin-degenerate ABS with positive energy $E_\uparrow=E_\downarrow=E$, plus the 
particle-hole partner states at energy $-E$.  In order to avoid the double-counting problem, one may employ the 
semiconductor picture \cite{Zazunov2014,Ackermann2023},  where
one retains only the single-particle ABSs with, say, energy $E_\uparrow$ and $-E_\downarrow$, see Fig.~\ref{fig2}(a).  The corresponding four many-body states
are shown in the left part of Fig.~\ref{fig2}(b); for details, see below.  Alternatively, in the excitation picture, 
 we instead retain the two positive energy levels $(E_\uparrow, E_\downarrow)$, where the corresponding many-body states are
 shown in the right part of Fig.~\ref{fig2}(b).  
 
 Let us summarize the many-body Andreev states for this example, as shown in Fig.~\ref{fig2}(b).
(i) In the semiconductor picture, the ground state $|g\rangle$ is obtained by filling the energy level $-E$ and leaving the energy level $+E$ empty.  In the excitation picture, both energy levels $E_{\sigma=\uparrow,\downarrow}$ are empty. We denote the ground state as $|00\rangle$ in the excitation picture.  This state has (by convention \cite{Ackermann2023}) 
even parity.
(ii) In the even parity sector, there is one excited state with excitation energy $2E$ above the ground state.
In the semiconductor picture, the lower level is empty but now the upper level is occupied.  This state
has been labelled $|e\rangle$ in Ref.~\cite{Ackermann2023}. In the 
excitation picture, both levels $E_\sigma$ are occupied, and the state is thus denoted as $|11\rangle$.
(iii) In the odd parity sector, there are two degenerate states with energy $E$ above the ground state.  In
the semiconductor picture, the state $|-\rangle$ has both levels $\pm E$ empty, while the state $|+\rangle$ has both levels occupied \cite{Ackermann2023}. In the excitation picture, one occupies only one of the two states. Here the two corresponding odd-parity states are called $|01\rangle$ and $|10\rangle$, respectively.

\begin{figure}
\begin{center}
    \includegraphics[width=0.499\textwidth]{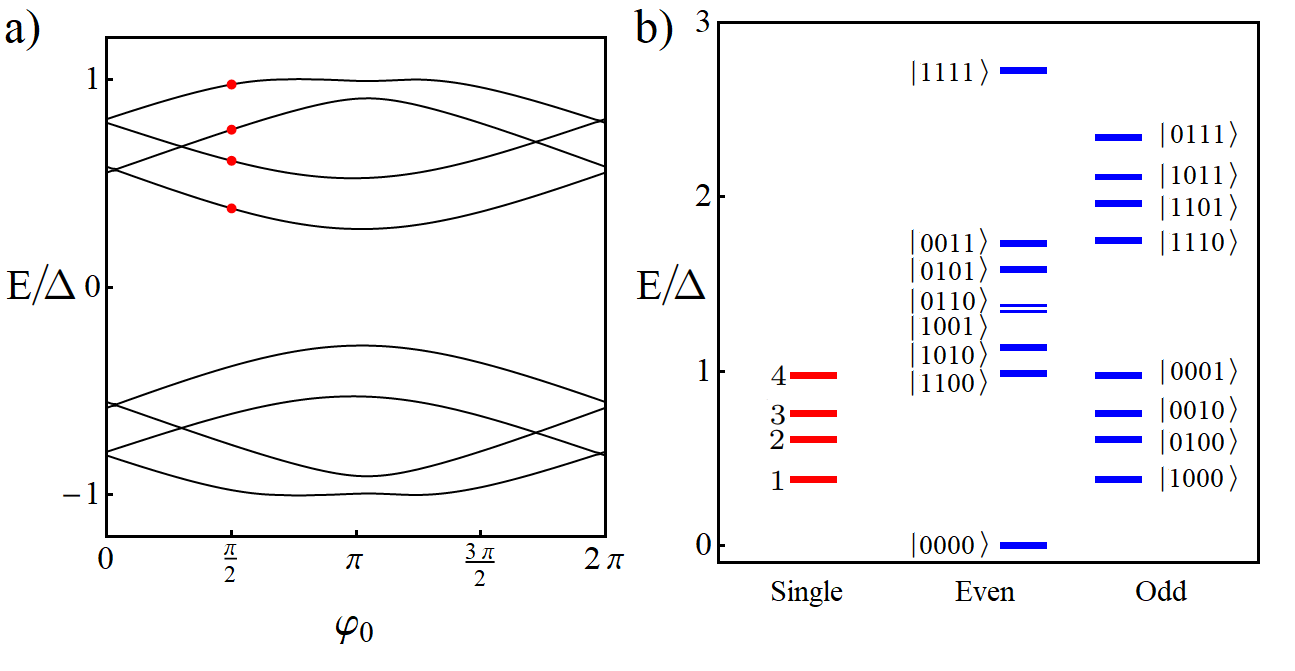}
    \caption{Andreev state dispersion and many-body Andreev states for a junction with SOI and Zeeman field.
    (a) ABS dispersion $E_\lambda$ vs phase difference $\varphi_0$ obtained numerically from Eq.~\eqref{detzero}.  The red dots indicate the four positive-energy states for $\varphi_0=\pi/2$. We here consider $L=1.5\xi_0, k_0\xi_0=0.1, {\cal T}=0.75, \gamma_{\rm SO}^{}=0.14$, $v_0=v_F$, and $\left|{\bf b}\right|=0.2 \Delta$ with $b_x/b_z=3$ and $b_y=0$. 
 (b) The red levels show the positive-energy single-particle ABS levels for $\varphi_0=\pi/2$, cf.~the red
      dots in panel (a). The sixteen possible many-body states are shown in the excitation picture as
      blue levels. We distinguish the even and odd fermion parity sectors. 
      The notation $|n_1n_2n_3n_4\rangle$ with $n_\lambda\in \{0,1\}$ means that the energy level $E_\lambda$ is either empty or occupied.  }
    \label{fig3}
\end{center}
\end{figure}

We now consider a junction in the presence of the SOI and the Zeeman field.  (In the absence of the Zeeman field, the Kramers degeneracy takes over the role of spin degeneracy.)  For instance, for a weak link of intermediate length $L\approx \xi_0=v_F/\Delta$, one typically finds four single-particle ABSs at positive energies, where two spin-degenerate levels split into four levels if both SOI and a Zeeman term are present.  An example is shown in Fig.~\ref{fig3}(a). In such cases, 
we find that the semiconductor picture is not useful for constructing a many-body formulation of the theory since it is
ambiguous how to select pairs of positive and negative energy states. 
From now on, we therefore use the excitation picture throughout. This picture allows us to directly circumvent double-counting 
problems in the many-body theory by construction.
For the case $L\approx \xi_0$, with fixed phase difference $\varphi_0$,  we order the positive ABS energies by increasing energy, $0 \leq E_1 \leq E_2 \leq E_3 \leq E_4 < \Delta$, see Fig.~\ref{fig3}(b). The resulting 16 many-body Andreev states are written 
as $|n_1 n_2 n_3 n_4\rangle$ with $n_\lambda\in \{0,1\}$, where $n_\lambda=0$ ($n_\lambda=1$) means that the energy level $E_\lambda$ is unoccupied (occupied). The ground state is then given by $|0000\rangle$. One can group those states into even- and odd-parity
states, see Fig.~\ref{fig3}(b).

\begin{figure}
\begin{center}
    \includegraphics[width=0.5\textwidth]{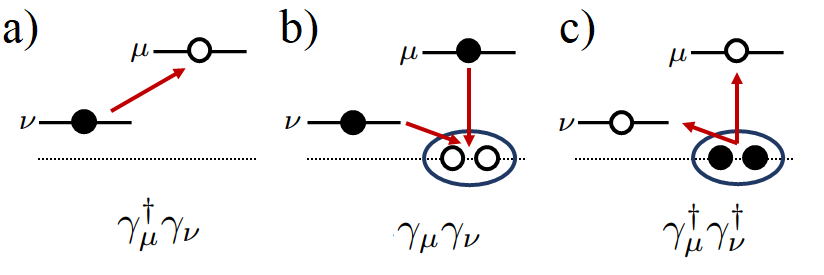}
    \caption{Illustration of the three contributions to the current matrix element \eqref{JC2} in the excitation picture. 
    (a) Transition between positive-energy BdG single-particle states $\nu\to \mu$, where
    the filled (empty) dot implies that the state $|\nu\rangle$ ($|\mu\rangle$) is initially filled (empty).
    The diagram in panel (b) [panel (c)] involves fermionic pair annihilation [creation] processes,
    along with the creation [annihilation] of a zero-energy Cooper pair. 
    Such processes emerge in the excitation picture due to negative-energy BdG states.
    }
    \label{fig4}
\end{center}
\end{figure}

 In the Schr\"odinger picture, the current operator \eqref{JC} then takes the form
\begin{equation}\label{JC2}
    \mathcal{I} = \sum_{\mu\ne \nu} \left( 2\mathcal{I}_{\mu, \nu}^{}\gamma^{\dagger}_{\mu} \gamma_{\nu}^{} +
     \mathcal{I}_{\bar\mu, \nu}^{}\gamma_{\mu}^{} \gamma_{\nu}^{}+
    \mathcal{I}_{\mu, \bar\nu}^{} \gamma^{\dagger}_{\mu} \gamma_{\nu}^{\dagger} 
    \right),
\end{equation}
where summations are taken over non-negative BdG energy solutions only. We  here
used the particle-hole relations $\gamma^{}_{\bar \nu}=\gamma^\dagger_\nu$ and $E_{\bar \nu}=-E_\nu$, which imply
 $\mathcal{I}_{\mu,\nu}=-\mathcal{I}_{\bar\nu,\bar\mu}$.   The term with $\mu=\nu$ has been excluded in 
Eq.~\eqref{JC2} since it does not contribute to the dynamical equations below. 
The possible  transitions contributing to the current matrix elements \eqref{JC2} are illustrated in Fig.~\ref{fig4}.
The first term in Eq.~\eqref{JC2} describes transitions between BdG single-particle eigenstates with
quantum numbers $\nu\to\mu$, see Fig.~\ref{fig4}(a). 
In the other two terms, we encounter fermionic pair annihilation or creation processes.  Such processes 
effectively arise from terms mixing ABSs with positive and negative energies in the excitation picture,
see also Ref.~\cite{Park2017}.  As shown in Fig.~\ref{fig4}(b,c), those processes involve the creation or annihilation of a 
Cooper pair, respectively.

\subsection{Lindblad equation}\label{sec3b}

We now turn to the dynamical equations for the density matrix $\rho(t)$ for the fermionic part of the system.  
Assuming that the harmonic oscillator bath representing $H_{\rm env}$ in Eq.~\eqref{fullH} remains in thermal
equilibrium at temperature $T_{\rm env}$ at all times, we assume for the total density operator $\rho_{\rm tot}(t)\approx \rho(t) \otimes \rho_{\rm env}$
in the interaction picture.  We make the standard Born-Markov assumptions of weak system-bath coupling and short bath memory time, which for our system 
are met for $T_{\rm env}\agt 10^{-2} \Delta$ and dimensionless system-bath coupling strength $\kappa_0\ll 1$ \cite{Ackermann2023}.
After tracing over the environmental modes, we obtain
\begin{eqnarray}\nonumber
    \partial_t \rho &=& \int_0^{\infty} d\tau\, \mathfrak{D}(\tau) \Big[\mathcal{I}(t-\tau) \rho(t) \mathcal{I}(t) - \mathcal{I}(t) \mathcal{I}(t-\tau) \rho(t)\Big]\\
    && \qquad  + \text{H.c.}, \label{eom1}
\end{eqnarray}
with a bath correlation function $\mathfrak{D}(\tau)$. Introducing real and imaginary parts in the frequency domain,
\begin{equation}
     \label{bathcorr}
    \int_0^{\infty}d\tau\, \mathfrak{D}(\tau)e^{i\omega\tau} = X(\omega)+iY(\omega), 
\end{equation}
the imaginary part $Y(\omega)$ is neglected below since it only weakly renormalizes the BdG quasiparticle energies. This causes the so-called Lamb shifts.  For the population dynamics studied in Sec.~\ref{sec3c} below within the Born approximation, 
such Lamb shifts are irrelevant. However, if one wishes to study quantum coherences encoded by the off-diagonal entries of the 
density operator, $Y(\omega)$ may have to be included \cite{breuer2007theory}.

In terms of the spectral density of a microwave-circuit environment with resonance frequency $\Omega$,
dimensionless coupling strength $\kappa_0$, and damping constant $\eta_d$ \cite{Nazarov2009}, 
\begin{equation}\label{specdens}
J(\omega) = \frac{\kappa^2_0\eta_d}{
\pi}\left(\frac{1}{(\omega-\Omega)^2+\frac{\eta^2_d}{2}} - \frac{1}{(\omega+\Omega)^2+\frac{\eta^2_d}{2}}\right) ,
\end{equation}
we obtain 
\begin{equation}\label{Xdef}
    X(\omega) = \pi J(\omega) [n_B(\omega)+1],  
\end{equation}
where $n_B(\omega) = (e^{\omega/T_{\text{env}}}-1)^{-1}$ is the Bose-Planck distribution.  
We also include a background Ohmic spectral density in $J(\omega)$, 
$J_{\text{ohm}} = 2\alpha_0 \omega e^{-|\omega|/\omega_c},$  with a dimensionless coupling $\alpha_0 \ll 1$ and the ultraviolet cutoff frequency $\omega_c$. 
In any case, the spectral density is defined to be asymmetric, $J(-\omega)=-J(\omega)$. 

For a given jump operator $c$, we employ the standard dissipator superoperator ${\cal L}[c]$   defined as \cite{breuer2007theory} 
\begin{equation}
    {\cal L}[c]\rho= c\rho c^\dagger - \frac12 \{ c^\dagger c,\rho\},
\end{equation}
where $\{\cdot,\cdot\}$ is the anticommutator. Inserting  $\mathcal{I}(t)$ obtained from Eq.~\eqref{JC2} into Eq.~\eqref{eom1}, and using $H_0$ in Eq.~\eqref{bdgf}, 
we then obtain a \emph{Lindblad master equation} \cite{breuer2007theory,Nava2021} for the time evolution of the fermionic density operator, 
\begin{eqnarray} \nonumber  
    \partial_t \rho &= &-i\sum_\nu E_\nu [\gamma_\nu^\dagger \gamma_\nu^{}, \rho(t)] +  \sum_{\mu,\nu} \Bigl( 
    \Gamma_{\mu,\nu} \,{\cal L}\left[\gamma_\mu^\dagger\gamma_\nu^{}\right]\rho(t)  + \\ \label{lindblad}
         &+& \frac12\left( \Gamma_{\bar\mu,\nu}\, {\cal L} \left[ \gamma_{\mu}^{ } \gamma_{\nu}^{ }\right] \rho(t)
         + \Gamma_{\mu,\bar\nu} \,{\cal L} \left[ \gamma_{\mu}^{\dagger} \gamma_{\nu}^{\dagger}\right] \rho(t)\right)
    \Bigr).
\end{eqnarray}
Since we work in the excitation picture, all summations over indices $\nu$ or $\mu$ involve only non-negative
(ABS or continuum) quasiparticle energies $E_\nu$ and $E_\mu$. Using Eq.~\eqref{Xdef},
the corresponding transition rates, see also Fig.~\ref{fig4}, are  given by
\begin{equation}
    \Gamma_{a,b}  =  2X(E_b-E_a)\, |\mathcal{I}_{a,b}|^2  \label{transrates},
\end{equation}
with the indices $a \in \{ \mu,\bar\mu\}$ and $b\in\{\nu,\bar\nu\}$, including both positive and negative BdG energy levels. 
For $a=\bar\mu$, we define $\bar a=\mu$.
(We recall our notation $\bar \nu$ for the particle-hole partner state with negative energy $E_{\bar \nu}=-E_\nu$ and quasiparticle
operator $\gamma_{\bar \nu}^{}=\gamma_\nu^\dagger$.) 

The transition rates in Eq.~\eqref{transrates} satisfy certain \emph{symmetry relations}. First, since the environment is in thermal equilibrium, we obtain the detailed balance relation 
\begin{equation}\label{detbal}
    \Gamma_{a,b}=e^{(E_b-E_a)/T_{\rm env}} \,\Gamma_{b,a}.
\end{equation} 
In addition, from the particle-hole symmetry in Eq.~\eqref{phs}, we infer the symmetry relation
\begin{equation}   \label{symrel}
    \Gamma_{a,b} = \Gamma_{\bar b,\bar a}.
\end{equation} 

In order to focus on the time evolution of the many-body Andreev states, we next trace over the quasiparticle continuum states.  
In general, this is a difficult task, and we here follow Ref.~\cite{Ackermann2023} by making two assumptions.  First, we assume 
that entanglement between the Andreev sector and the continuum sector can be neglected at all times such that the fermionic density
operator factorizes, $\rho(t)\approx \rho_A(t)\otimes \rho_c(t)$.  Here, $\rho_A(t)$ is the reduced density operator
of the Andreev sector while $\rho_c(t)$ describes the continuum quasiparticle sector.  Second, we  
assume that $\rho_c(t)$ can be written in terms of an equilibrium distribution function, $\tilde{n}_p$, which depends only on the continuum-state quantum numbers $p=(E,s,\sigma)$,
\begin{equation}
     \rho_{\text{c}}(t) = \prod_p \left( \tilde{n}_p| 1_p \rangle \langle 1_p| + (1-\tilde{n}_p) |0_p \rangle \langle 0_p| \right),
\end{equation}
where $|1_p\rangle=\gamma_p^\dagger|0_p\rangle$ and $|0_p\rangle$ are the eigenstates of $\gamma_p^\dagger \gamma_p^{}$ with eigenvalue $1$ and $0$, respectively.
Note that the product extends only over $E>\Delta$ solutions since we work in the excitation picture.
For the distribution function, we choose a Fermi-Dirac distribution parametrized by a ``quasiparticle temperature'' $T_{\rm qp}$,
\begin{equation}\label{fdqp}
    \tilde{n}_p =  (e^{E/T_{\rm qp}}+1)^{-1}.
\end{equation}
We note that $\tilde{n}_{\bar p}=1-\tilde{n}_p$ holds for the corresponding negative-energy state.
The temperature $T_{\rm qp}$ may differ from the temperature $T_{\rm env}$ of the electromagnetic environment.  For instance,
in order to describe quasiparticle poisoning effects due to the presence of excess above-gap quasiparticles, at least in a qualitative manner,  we consider $T_{\rm qp}>T_{\rm env}$.

\begin{figure}
\begin{center}
    \includegraphics[width=0.46\textwidth]{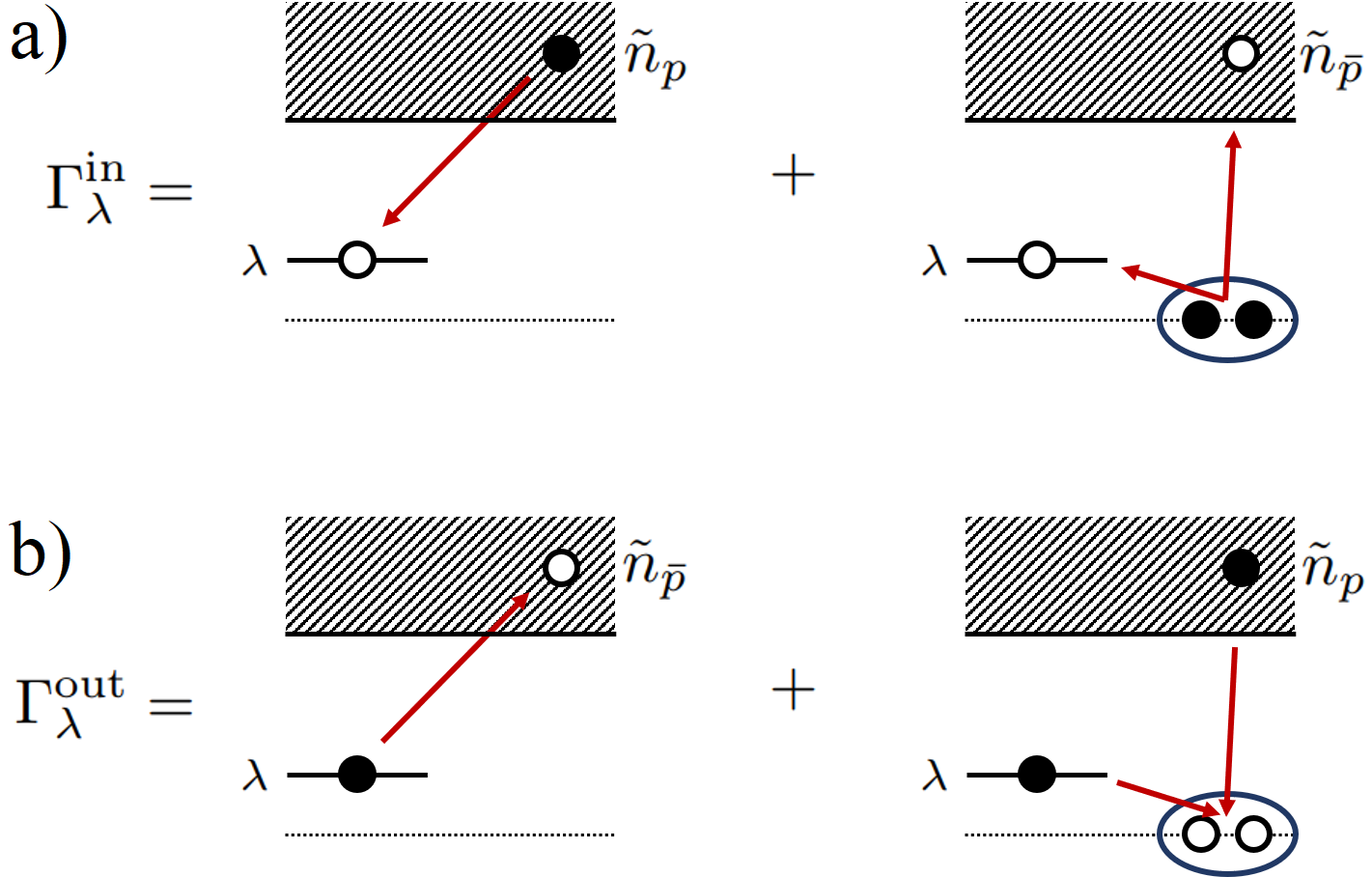}
    \caption{Illustration of the transition rates $\Gamma_\lambda^{\rm in}$  [panel (a)] and $\Gamma_\lambda^{\rm out}$ [panel (b)] 
    connecting the ABS level $\lambda$ to the quasiparticle continuum, see Eq.~\eqref{transrates}.  The distribution function $\tilde n_p$  of the continuum particles is defined in Eq.~\eqref{fdqp}. Note that in the excitation picture, ``anomalous'' processes involving a Cooper pair have to be taken into account.
    }
    \label{fig5}
\end{center}
\end{figure}

Using the symmetry relation \eqref{symrel} and performing the trace over the continuum sector in Eq.~\eqref{lindblad}, we finally arrive at a Lindblad equation describing only the Andreev sector, 
\begin{eqnarray}  \nonumber
\partial_t \rho_A &= &-i\sum_\lambda E_\lambda [\gamma_\lambda^\dagger \gamma_\lambda^{}, \rho_A] +  \sum_{\lambda,\lambda'} \Biggl( 
\Gamma_{\lambda,\lambda'} \,{\cal L}\left[\gamma_\lambda^\dagger\gamma_{\lambda'}^{}\right]\rho_A   +  \\ &+&
\frac12\left(\Gamma_{\bar\lambda,\lambda'}\, {\cal L} \left[ \gamma_{\lambda}^{ } \gamma_{\lambda'}^{ }\right] \rho_A
+ \Gamma_{\lambda,\bar\lambda'} \,{\cal L} \left[ \gamma_{\lambda}^{\dagger} \gamma_{\lambda'}^{\dagger}\right] \rho_A\right)
    \Biggr) \nonumber \\ \label{lindbladA}
 &+& \sum_{\lambda} \left(\Gamma_\lambda^{\rm in} \, \mathcal{L}[\gamma^{\dagger}_{\lambda}] \rho_A + 
  \Gamma_\lambda^{\rm out} \, \mathcal{L}[\gamma^{}_{\lambda}] \rho_A  \right).
\end{eqnarray}
Apart from the transition rates \eqref{transrates} between ABSs, Eq.~\eqref{lindbladA} also involves transition rates connecting the sub-gap Andreev and the above-gap continuum sector, 
\begin{eqnarray} \label{transrateinout}
  \Gamma_{\lambda}^{\text{in}} &=&  \sum_p \left( \Gamma_{ \lambda,p} \tilde{n}_p + \Gamma_{\lambda,\bar p} \tilde{n}_{\bar p}  \right ), \\
  \Gamma_{\lambda}^{\text{out}}  &=& \sum_p\left( \Gamma_{p,\lambda}\tilde{n}_{\bar p} + \Gamma_{\bar p, \lambda} \tilde{n}_{p} \right)= \Gamma_{\bar\lambda}^{\text{in}}.\nonumber
\end{eqnarray}
We emphasize again that summations over ABS indices $\lambda$ and over continuum indices $p$ involve only non-negative energy levels. 
The corresponding processes are schematically illustrated in Fig.~\ref{fig5}.

\begin{table*}
\begin{center}
\begin{tabular}{|c || c| c| c| c|c|c|c|c|c|c|c|c|c|c|c|c|} 
 \hline
  & $\ket{0000}$ & $\ket{1000}$ & $\ket{0010}$ & $\ket{1010}$ & $\ket{0101}$ & $\ket{1111}$ & $\ket{1100}$ & $\ket{0011}$ & $\ket{1001}$ & $\ket{0110}$ & $\ket{0100}$ &$\ket{0001}$ & $\ket{1110}$ & $\ket{1011}$ & $\ket{1101}$ & $\ket{0111}$ \\
 \hline\hline 
 $\bra{0000}$ &   & $\Gamma_{\bar{1}}^{\text{in}}$ & $\Gamma_{\bar{3}}^{\text{in}}$ & $\Gamma_{\bar{1},3}$ & $\Gamma_{\bar{2},4}$ & 0 & $\Gamma_{\bar{1},2}$ & $\Gamma_{\bar{3},4}$ & 
 $\Gamma_{\bar{1},4}$ & $\Gamma_{\bar{2},3}$ &  $\Gamma_{\bar{2}}^{\text{in}}$ & $\Gamma_{\bar{4}}^{\text{in}}$ & 0 & 0 & 0 & 0\\[0.5ex]  
 $\bra{1000}$ & $\Gamma_{1}^{\text{in}}$ &  & $\Gamma_{1,3}$ & $\Gamma_{\bar{3}}^{\text{in}}$ & 0 & 0 & $\Gamma_{\bar{2}}^{\text{in}}$ & 0 & $\Gamma^{\text{in}}_{\bar{4}}$ & 0 & $\Gamma_{1,2}$ & $\Gamma_{1,4}$ & $\Gamma_{\bar{2},3}$ & $\Gamma_{\bar{3},4}$ & $\Gamma_{\bar{2},4}$ & 0\\[0.5ex]
 $\bra{0010}$ & $\Gamma^{\text{in}}_{3}$ & $\Gamma_{3,1}$ &   & $\Gamma^{\text{in}}_{\bar{1}}$ & 0 & 0 & 0 & $\Gamma^{\text{in}}_{\bar{4}}$ & 0 & $\Gamma^{\text{in}}_{\bar{2}}$ & $\Gamma_{3,2}$ & $\Gamma_{3,4}$ & $\Gamma_{\bar{1},2}$ & $\Gamma_{\bar{1},4}$ & 0 & $\Gamma_{\bar{2},4}$ \\[0.5ex] 
 $\bra{1010}$ & $\Gamma_{1,\bar{3}}$ & $\Gamma^{\text{in}}_{3}$ & $\Gamma^{\text{in}}_{1}$ &  & 0 & $\Gamma_{\bar{2},4}$ & $\Gamma_{3,2}$ & $\Gamma_{1,4}$ &
 $\Gamma_{3,4}$ & $\Gamma_{1,2}$ & 0 & 0 & $\Gamma^{\text{in}}_{\bar{2}}$ & $\Gamma^{\text{in}}_{\bar{4}}$ & 0 & 0\\[0.5ex]
 $\bra{0101}$ & $\Gamma_{2,\bar{4}}$ & 0 & 0 & 0 &   & $\Gamma_{\bar{1},3}$ & $\Gamma_{4,1}$ & $\Gamma_{2,3}$ &
 $\Gamma_{2,1}$ & $\Gamma_{4,3}$ & $\Gamma^{\text{in}}_{4}$ & $\Gamma^{\text{in}}_{2}$ & 0 & 0 & $\Gamma^{\text{in}}_{\bar{1}}$ & $\Gamma^{\text{in}}_{\bar{3}}$ \\[0.5ex]
 $\bra{1111}$ & 0 & 0 & 0 & $\Gamma_{2,\bar{4}}$ & $\Gamma_{1,\bar{3}}$ &   & $\Gamma_{3,\bar{4}}$ & $\Gamma_{1,\bar{2}}$ &
 $\Gamma_{2,\bar{3}}$ & $\Gamma_{1,\bar{4}}$ & 0 & 0 & $\Gamma^{\text{in}}_{4}$ & $\Gamma^{\text{in}}_{2}$ &$\Gamma^{\text{in}}_{3}$ & $\Gamma^{\text{in}}_{1}$ \\[0.5ex] 
 $\bra{1100}$ & $\Gamma_{1,\bar{2}}$ & $\Gamma^{\text{in}}_{2}$ & 0 & $\Gamma_{2,3}$ & $\Gamma_{1,4}$ & $\Gamma_{\bar{3},4}$ &  & 0 & $\Gamma_{2,4}$ & $\Gamma_{1,3}$ & $\Gamma^{\text{in}}_{1}$ & 0 & $\Gamma^{\text{in}}_{\bar{3}}$ & 0 & $\Gamma^{\text{in}}_{\bar{4}}$ & 0\\[0.5ex] 
 $\bra{0011}$ & $\Gamma_{3,\bar{4}}$ & 0 & $\Gamma^{\text{in}}_{4}$ & $\Gamma_{4,1}$ & $\Gamma_{3,2}$ & $\Gamma_{\bar{1},2}$ & 0 & 
  & $\Gamma_{3,1}$ & $\Gamma_{4,2}$ & 0 & $\Gamma^{\text{in}}_{3}$ & 0 & $\Gamma^{\text{in}}_{\bar{1}}$ & 0 & $\Gamma^{\text{in}}_{\bar{2}}$ \\[0.5ex] 
 $\bra{1001}$ & $\Gamma_{1,\bar{4}}$ & $\Gamma^{\text{in}}_{4}$ & 0 & $\Gamma_{4,3}$ & $\Gamma_{1,2}$ & $\Gamma_{\bar{2},3}$ & $\Gamma_{4,2}$ & $\Gamma_{1,3}$ &  & 0 & 0 & $\Gamma^{\text{in}}_{1}$ & 0 & $\Gamma^{\text{in}}_{\bar{3}}$ & $\Gamma^{\text{in}}_{\bar{2}}$ & 0\\[0.5ex] 
 $\bra{0110}$ & $\Gamma_{2,\bar{3}}$ & 0 & $\Gamma^{\text{in}}_{2}$ & $\Gamma_{2,1}$ & $\Gamma_{3,4}$ & $\Gamma_{\bar{1},4}$ & 
 $\Gamma_{3,1}$ & $\Gamma_{2,4}$ & 0 &   & $\Gamma^{\text{in}}_{3}$ & 0 & $\Gamma^{\text{in}}_{\bar{1}}$ & 0 & 0 & $\Gamma^{\text{in}}_{\bar{4}}$ \\[0.5ex]
 $\bra{0100}$ & $\Gamma^{\text{in}}_{2}$  & $\Gamma_{2,1}$ & $\Gamma_{2,3}$ & 0 & $\Gamma^{\text{in}}_{\bar{4}}$ & 0 & $\Gamma^{\text{in}}_{\bar{1}}$ & 0 & 0 & $\Gamma^{\text{in}}_{\bar{3}}$ &   & $\Gamma_{2,4}$ & $\Gamma_{\bar{1},3}$ & 0 & $\Gamma_{\bar{1},4}$ & $\Gamma_{\bar{3},4}$  \\[0.5ex] 
 $\bra{0001}$ & $\Gamma^{\text{in}}_{4}$  & $\Gamma_{4,1}$ & $\Gamma_{3,4}$ & 0 & $\Gamma^{\text{in}}_{\bar{2}}$ & 0 & 0 & $\Gamma^{\text{in}}_{\bar{3}}$ & $\Gamma^{\text{in}}_{\bar{1}}$ &0 & 
 $\Gamma_{4,2}$ &   & 0 & $\Gamma_{\bar{1},3}$ & $\Gamma_{\bar{1},2}$ & $\Gamma_{\bar{2},3}$   \\[0.5ex]  
 $\bra{1110}$ & 0 & $\Gamma_{2,\bar{3}}$ & $\Gamma_{1,\bar{2}}$ & $\Gamma^{\text{in}}_{2}$ & 0 & $\Gamma^{\text{in}}_{\bar{4}}$ &
 $\Gamma^{\text{in}}_{3}$ & 0 & 0 & $\Gamma^{\text{in}}_{1}$ & $\Gamma_{1,\bar{3}}$ & 0 & & $\Gamma_{2,4}$ & $\Gamma_{3,4}$ & $\Gamma_{1,4}$ \\[0.5ex]
 $\bra{1011}$ & 0 & $\Gamma_{3,\bar{4}}$ & $\Gamma_{1,\bar{4}}$ &  $\Gamma^{\text{in}}_{4}$ & 0 & $\Gamma^{\text{in}}_{\bar{3}}$ & 0 & $\Gamma^{\text{in}}_{1}$ & $\Gamma^{\text{in}}_{3}$ & 0 & 0 &  $\Gamma_{1,\bar{3}}$ & $\Gamma_{4,2}$ &   &  $\Gamma_{3,2}$ & $\Gamma_{1,2}$  \\[0.5ex]  
 $\bra{1101}$ & 0 & $\Gamma_{2,\bar{4}}$ & 0 & 0 & $\Gamma^{\text{in}}_{1}$ & $\Gamma^{\text{in}}_{\bar{3}}$ &
 $\Gamma^{\text{in}}_{4}$ & 0 & $\Gamma^{\text{in}}_{2}$ & 0 & $\Gamma_{1,\bar{4}}$ & $\Gamma_{1,\bar{2}}$ & $\Gamma_{4,3}$ & $\Gamma_{2,3}$ & & $\Gamma_{1,3}$ \\[0.5ex] 
 $\bra{0111}$ & 0 & 0 & $\Gamma_{2,\bar{4}}$  & 0 & $\Gamma^{\text{in}}_{3}$ & $\Gamma^{\text{in}}_{\bar{1}}$ & 0 & $\Gamma^{\text{in}}_{2}$ & 0 & $\Gamma^{\text{in}}_{4}$ & 
 $\Gamma_{3,\bar{4}}$ & $\Gamma_{2,\bar{3}}$ & $\Gamma_{4,1}$ & $\Gamma_{2,1}$ & $\Gamma_{3,1}$ &  \\[0.5ex]  
 \hline
\end{tabular}
\caption{Off-diagonal matrix elements $M_{{\bf n},{\bf n'}}$ for the matrix $\mathbf{M}$ in Eq.~(\ref{matrixrate}), expressed in terms of the transition rates \eqref{transrates} and \eqref{transrateinout}.
Here $|{\bf n}\rangle=|n_1n_2n_3n_4\rangle$ with $n_\lambda\in \{0,1\}$ labels the 16 possible many-body Andreev states.
The indices $\lambda\in \{1,2,3,4\}$ refer to the four single-particle ABS states, see, e.g., Fig.~\ref{fig3}, where $\bar{\lambda}$ corresponds to the negative-partner state. Diagonal matrix elements $M_{{\bf n},{\bf n}}$ are given in Table \ref{tab2}.
}
\label{matrix} 
\end{center}
\end{table*}

\begin{table*}
\begin{center}
\begin{tabular}{|c || c |} 
 \hline
 $|{\bf n}\rangle$ & $- M_{{\bf n},{\bf n}}$ \\
 \hline\hline
 $\ket{0000}$ & $\Gamma_{1,\bar{2}} + \Gamma_{1,\bar{3}} + \Gamma_{2,\bar{3}}+ \Gamma_{1,\bar{4}} + \Gamma_{2,\bar{4}} + \Gamma_{3,\bar{4}}   + \Gamma^{\text{in}}_1 + \Gamma^{\text{in}}_2 + \Gamma^{\text{in}}_3 + \Gamma^{\text{in}}_4$  \\[0.5ex]  
 $\ket{1000}$ & $\Gamma_{2,1} + \Gamma_{3,1} + \Gamma_{4,1} +  \Gamma_{2,\bar{3}} + \Gamma_{2,\bar{4}} + \Gamma_{3,\bar{4}}
+ \Gamma_2^{\text{in}}+\Gamma_3^{\text{in}}+\Gamma_4^{\text{in}} + \Gamma_{\bar{1}}^{\text{in}} $\\[0.5ex]
 $\ket{0010}$ & $\Gamma_{1,3} + \Gamma_{2,3} + \Gamma_{4,3} + \Gamma_{1,\bar{2}} + \Gamma_{1,\bar{4}} + \Gamma_{2,\bar{4}}  
+ \Gamma_1^{\text{in}}+\Gamma_2^{\text{in}}+\Gamma_4^{\text{in}} + \Gamma_{\bar{3}}^{\text{in}}$ \\[0.5ex] 
 $\ket{1010}$ & $\Gamma_{2,1} + \Gamma_{4,1} + \Gamma_{2,3} + \Gamma_{4,3} + \Gamma_{2,\bar{4}} + \Gamma_{\bar{1},3} 
    + \Gamma_2^{\text{in}}+\Gamma_4^{\text{in}}+\Gamma_{\bar{1}}^{\text{in}} + \Gamma_{\bar{3}}^{\text{in}}$ \\[0.5ex]
 $\ket{0101}$ & $\Gamma_{1,2} + \Gamma_{3,2} + \Gamma_{1,4} + \Gamma_{3,4} + \Gamma_{1,\bar{3}} + \Gamma_{\bar{2},4}
    + \Gamma_1^{\text{in}}+\Gamma_3^{\text{in}}+\Gamma_{\bar{2}}^{\text{in}} + \Gamma_{\bar{4}}^{\text{in}}$   \\[0.5ex]
$\ket{1111}$& $\Gamma_{\bar{1},2} + \Gamma_{\bar{1},3} + \Gamma_{\bar{2},3}+ \Gamma_{\bar{1},4} + \Gamma_{\bar{2},4} + \Gamma_{\bar{3},4} + \Gamma^{\text{in}}_{\bar 1} + \Gamma^{\text{in}}_{\bar 2} + \Gamma^{\text{in}}_{\bar 3} + \Gamma^{\text{in}}_{\bar 4}$ \\[0.5ex]
$\ket{1100}$ & $\Gamma_{3,1} + \Gamma_{4,1} + \Gamma_{3,2} + \Gamma_{4,2} + \Gamma_{3,\bar{4}} + \Gamma_{\bar{1},2} 
     + \Gamma_3^{\text{in}}+\Gamma_4^{\text{in}}+\Gamma_{\bar 1}^{\text{in}} + \Gamma_{\bar 2}^{\text{in}}$ \\[0.5ex]
     $\ket{0011}$ & $\Gamma_{1,3} + \Gamma_{1,4} + \Gamma_{2,3} + \Gamma_{2,4} + \Gamma_{1,\bar{2}} + \Gamma_{\bar{3},4} 
     + \Gamma_1^{\text{in}}+\Gamma_2^{\text{in}}+\Gamma_{\bar 3}^{\text{in}} + \Gamma_{\bar 4}^{\text{in}}$ \\[0.5ex] 
    $\ket{1001}$  & $\Gamma_{2,1} + \Gamma_{3,1} + \Gamma_{2,4} + \Gamma_{3,4} + \Gamma_{2,\bar{3}} + \Gamma_{\bar{1},4}
     + \Gamma_2^{\text{in}}+\Gamma_3^{\text{in}}+\Gamma_{\bar 1}^{\text{in}} + \Gamma_{\bar 4}^{\text{in}}$ \\[0.5ex]
    $\ket{0110}$ &$\Gamma_{1,2} + \Gamma_{4,2} + \Gamma_{1,3} + \Gamma_{4,3} + \Gamma_{1,\bar{4}} + \Gamma_{\bar{2},3}
     + \Gamma_1^{\text{in}}+\Gamma_4^{\text{in}}+\Gamma_{\bar{2}}^{\text{in}} + \Gamma_{\bar 3}^{\text{in}}$ \\[0.5ex]
    $\ket{0100}$ & $\Gamma_{1,2} + \Gamma_{3,2} + \Gamma_{4,2} + \Gamma_{1,\bar{3}} + \Gamma_{1,\bar{4}} + \Gamma_{3,\bar{4}}    + \Gamma_1^{\text{in}}+\Gamma_3^{\text{in}}+\Gamma_4^{\text{in}} + \Gamma_{\bar 2}^{\text{in}}$\\[0.5ex]
    $\ket{0001}$ & $\Gamma_{1,4} + \Gamma_{2,4} + \Gamma_{3,4} + \Gamma_{1,\bar{2}} + \Gamma_{1,\bar{3}} + \Gamma_{2,\bar{3}}     + \Gamma_1^{\text{in}}+\Gamma_2^{\text{in}}+\Gamma_3^{\text{in}} + \Gamma_{\bar 4}^{\text{in}}$ \\[0.5ex]
    $\ket{1110}$ & $\Gamma_{4,1} + \Gamma_{4,2} + \Gamma_{4,3} + \Gamma_{\bar{1},2} + \Gamma_{\bar{1},3} + \Gamma_{\bar{2},3}
     + \Gamma_4^{\text{in}}+\Gamma_{\bar{1}}^{\text{in}}+\Gamma_{\bar{2}}^{\text{in}} + \Gamma_{\bar 3}^{\text{in}}$ \\[0.5ex]
    $\ket{1011}$ & $\Gamma_{2,1} + \Gamma_{2,3} + \Gamma_{2,4} + \Gamma_{\bar{1},3} + \Gamma_{\bar{1},4} + \Gamma_{\bar{3},4}
    + \Gamma_2^{\text{in}}+\Gamma_{\bar{1}}^{\text{in}}+\Gamma_{\bar{3}}^{\text{in}} + \Gamma_{\bar{4}}^{\text{in}}$ \\[0.5ex]
    $\ket{1101}$ & $\Gamma_{3,1} + \Gamma_{3,2} + \Gamma_{3,4} + \Gamma_{\bar{1},2} + \Gamma_{\bar{1},4} + \Gamma_{\bar{2},4} 
     + \Gamma_3^{\text{in}}+\Gamma_{\bar{1}}^{\text{in}}+\Gamma_{\bar{2}}^{\text{in}} + \Gamma_{\bar{4}}^{\text{in}}$ \\[0.5ex]
    $\ket{0111}$ & $\Gamma_{1,2} + \Gamma_{1,3} + \Gamma_{1,4} + \Gamma_{\bar{2},3} + \Gamma_{\bar{2},4} + \Gamma_{\bar{3},4}
     + \Gamma_1^{\text{in}}+\Gamma_{\bar{2}}^{\text{in}}+\Gamma_{\bar{3}}^{\text{in}} + \Gamma_{ \bar{4} }^{\text{in}}$ \\[0.5ex]
 \hline 
\end{tabular}
\caption{Diagonal matrix elements $M_{{\bf n},{\bf n}}$ in Eq.~(\ref{matrixrate}) expressed in terms of the transition rates \eqref{transrates} and \eqref{transrateinout},
where $|{\bf n}\rangle=|n_1n_2n_3n_4\rangle$  labels the  many-body Andreev states.
Note that we specify $-M_{{\bf n},{\bf n}}$. Off-diagonal matrix elements are specified in Table \ref{matrix}.
}
\label{tab2} 
\end{center}
\end{table*}

\subsection{Population dynamics of many-body Andreev states}\label{sec3c}

As final step, we project the Lindblad equation \eqref{lindbladA} for the density operator $\rho_A(t)$ into the many-body Andreev states $|{\bf n}\rangle$. For clarity, we focus on cases with four spin-split positive-energy ABS solutions but for other cases one can proceed analogously. For the example in Fig.~\ref{fig3}, we have $|{\bf n }\rangle=|n_1n_2n_3n_4\rangle$,
where $n_\lambda\in \{0,1\}$ specifies whether the (non-negative) ABS level $E_\lambda$ is unoccupied or occupied.
 Physically, this corresponds to the case of intermediate-length junctions with $L\approx \xi_0$. The diagonal elements of $\rho_A$ represent the occupation probabilities of the respective 16 Andreev many-body states,
\begin{equation}
P_{\bf n}(t) = \langle {\bf n}| \rho_A(t) | {\bf n}\rangle, \quad \sum_{\bf n} P_{\bf n}(t)=1.
\end{equation} 
We combine these probabilities into a 16-dimensional vector $\mathbf{P}(t)$.  
Since the dynamics of  $\mathbf{P}(t)$ decouples from the off-diagonal part of $\rho_A(t)$, the occupation probabilities evolve
independently from quantum coherences.  In what follows, we then focus on the time evolution of $\mathbf{P}(t)$.  

From Eq.~\eqref{lindbladA}, by taking the appropriate matrix elements, we obtain a matrix rate equation of the form
\begin{align} \label{matrixrate}
     \mathbf{\dot P}(t) = \mathbf{M}\,\mathbf{P}(t),
\end{align}
where the $16\times 16$ matrix $\mathbf{M}$ is specified  in Tables \ref{matrix} and \ref{tab2}.
For the stationary state reached at asymptotically long times, $\dot{\mathbf{P}}_{\rm stat} = 0$, the steady-state occupation probabilities 
$\mathbf{P}_{\rm stat}$ follow by determining the kernel of the matrix ${\bf M}$.  In general, given the (real-valued and non-positive) eigenvalues $\lambda_k$ and the corresponding right eigenvectors ${\bf P}_k$ of ${\bf M}$,
 the general solution of Eq.~\eqref{matrixrate} follows as
 \begin{equation}\label{gensol}
     {\bf P}(t)= \sum_{k=1}^{16} c_k {\bf P}_k e^{\lambda_k t}.
 \end{equation}
 The coefficients $c_k$ are determined by matching Eq.~\eqref{gensol} to the initial configuration ${\bf P}(0)$ at time $t=0$.  By collecting the contributions from 
 the odd- and even-parity states only, we can define the probabilities $P_{\rm odd}(t)$ and $P_{\rm even}(t)=1-P_{\rm odd}(t)$ for occupying the respective
 parity sector.
 
\section{Results for the population dynamics}\label{sec4}

We now turn to a discussion of results obtained from Eq.~\eqref{matrixrate}.  For concreteness, we focus throughout on a weak link of intermediate length, $L=1.5\xi_0$, with contact transparencies
${\cal T}_1={\cal T}_2={\cal T}=0.75$.  One then generically finds
four positive-energy single-particle ABS solutions for given phase difference $\varphi_0$.
For the SOI parameter, we assume $\gamma_{\rm SO}^{}=0.14$ following the estimates in Ref.~\cite{Tosi2019}, but we also contrast our results to the case without SOI.  
Similarly, if the Zeeman field is switched on, we assume $|{\bf b}|=0.2\Delta$. For instance, taking Nb as superconductor and InAs as nanowire material, accounting for the rather large $g$-factors in such nanowires \cite{Bjoerk2005}, 
$|{\bf b}|=0.2\Delta$ translates into a field strength $\approx 0.2$~Tesla at low temperatures. We
then consider arbitrary directions of the field with respect to the polar axis defined by the SOI (which is the $z$-axis).  

\begin{figure}
\begin{center}
    \includegraphics[width=0.48\textwidth]{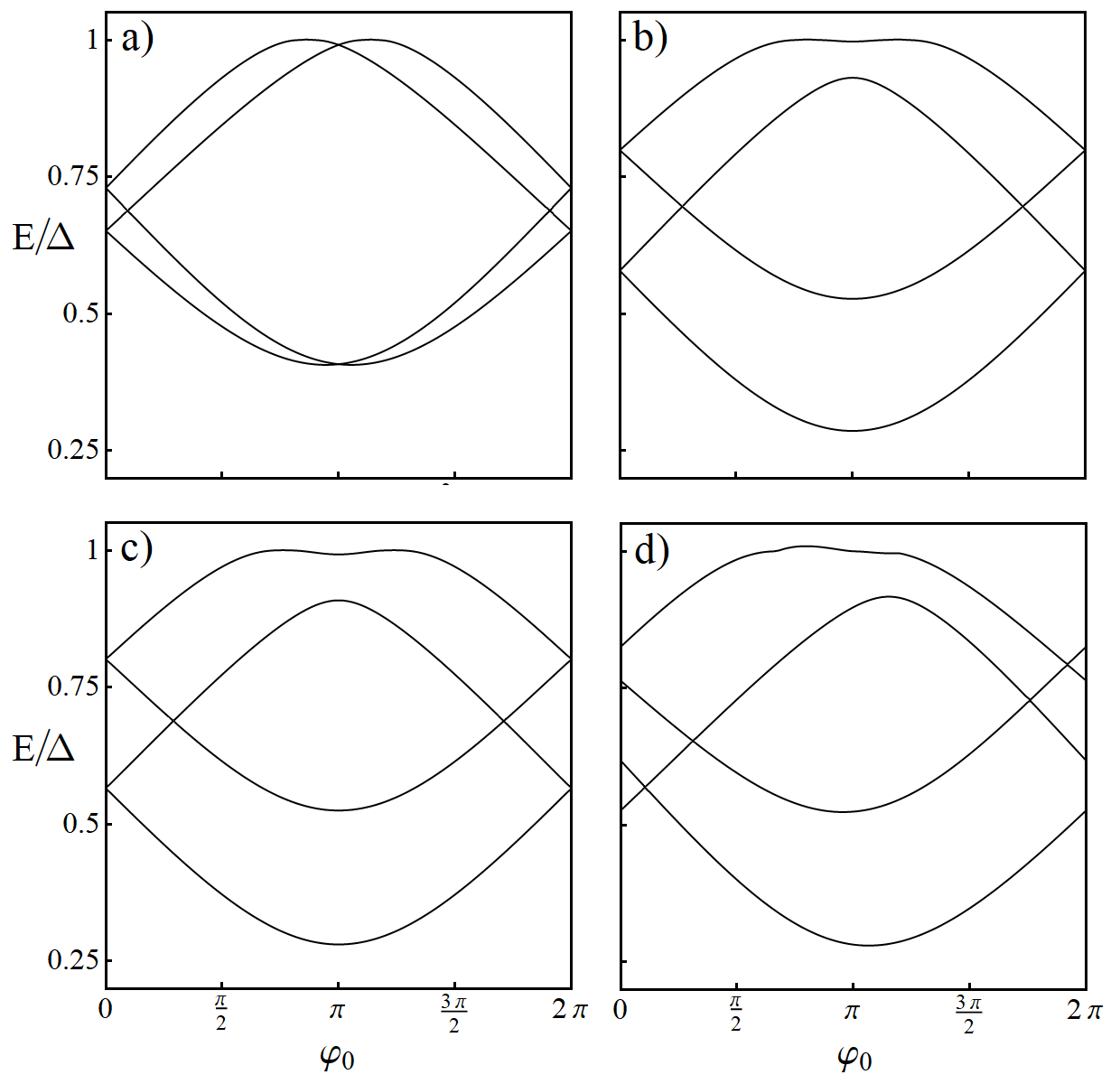}
    \caption{ABS dispersion $E_\lambda$ (with $E_\lambda>0$) vs $\varphi_0$ for a weak link with 
    $L=1.5\xi_0, k_0\xi_0=0.1, {\cal T}=0.75$, and $v_0=v_F$.   (a) Finite SOI strength 
    $\gamma^{}_{\rm SO}=0.14$ but vanishing Zeeman field, ${\bf b}=0$. (b) Vanishing SOI strength, $\gamma_{\rm SO}^{}=0$, with ${\bf b}$ along an arbitrary direction for $|{\bf b}|=0.2\Delta$. 
(c) Case $\gamma_{\rm SO}^{}=0.14$ and Zeeman field ${\bf b}$ along the $x$-direction with $|{\bf b}|=0.2\Delta$. (d) Same as in panel 
(c) but with the Zeeman field along the $z$-direction. }
    \label{fig6}
\end{center}
\end{figure}

One example for the single-particle ABS dispersion of such a junction has already been shown in Fig.~\ref{fig3}(a). 
In Fig.~\ref{fig6}, we show four additional examples, obtained by numerically solving Eq.~\eqref{detzero}.  In Fig.~\ref{fig6}(a), we observe that in the absence of the magnetic field, the Kramers
degeneracy at $\varphi_0=0,\pi$ takes over the role of the usual spin degeneracy. Moreover, the dispersion is 
symmetric, $E_\lambda(2\pi-\varphi_0)=E_\lambda(\varphi_0)$.  As seen in Fig.~\ref{fig6}(b), this symmetry of the dispersion is also found when we switch off the SOI but switch on the magnetic field.  However, there are no time-reversal-invariant points anymore.  In Fig.~\ref{fig6}(c), we consider the case where both SOI and Zeeman field are present, with the Zeeman field along the nanowire axis. Now all degeneracies are broken but the above symmetry still remains intact. 
While the spectrum looks very similar to the one in panel (b), there are small differences.  We note that the level crossings 
are not avoided crossings.
However, if the magnetic field is oriented along the polar axis of the SOI, we find   $E_\lambda(2\pi-\varphi_0)\ne E_\lambda(\varphi_0)$, as shown in Fig.~\ref{fig6}(d).
Incidentally, in such  cases, the anomalous Josephson effect and the superconducting diode effect will arise, see, e.g., Refs.~\cite{Zazunov2009,Brunetti2013,Zazunov2024}.

We now turn to the population dynamics $P_{\bf n}(t)$ of the respective many-body Andreev states $|{\bf n}\rangle$.  
We here assume that at times $t<0$, for a given parameter set, the system has been prepared in its steady state 
with probabilities ${\bf P}_{\rm stat}$, see Sec.~\ref{sec3c}.  At time $t=0$, one applies a short and strong microwave pulse of frequency $\Omega_d$. We assume that $\Omega_d$ is
resonant with a transition from the ground state $|{\bf n}_0\rangle=|0000\rangle$ to an excited many-body Andreev state $|{\bf n}\rangle=|n_1n_2n_3n_4\rangle$ of the same fermion parity,
i.e., $(-1)^{n_1+n_2+n_3+n_4}=+1$.  (The microwave drive cannot change the fermion parity.)
If the respective transition rate $M_{{\bf n},{\bf n}_0}$ in Table \ref{matrix} is finite (this condition
imposes a selection rule), population inversion between $|{\bf n}_0\rangle$ and $|{\bf n}\rangle$ can
be induced by the microwave pulse, as explained in Ref.~\cite{Ackermann2023}.  In this way, one can effectively 
study the effect of the microwave drive through a nonequilibrium initial condition in Eq.~\eqref{matrixrate}, 
where the occupation probabilities for the two levels $|{\bf n}_0\rangle$
and $|{\bf n}\rangle$ are exchanged with respect to their steady-state values.

\begin{figure}
\begin{center}
    \includegraphics[width=0.499\textwidth]{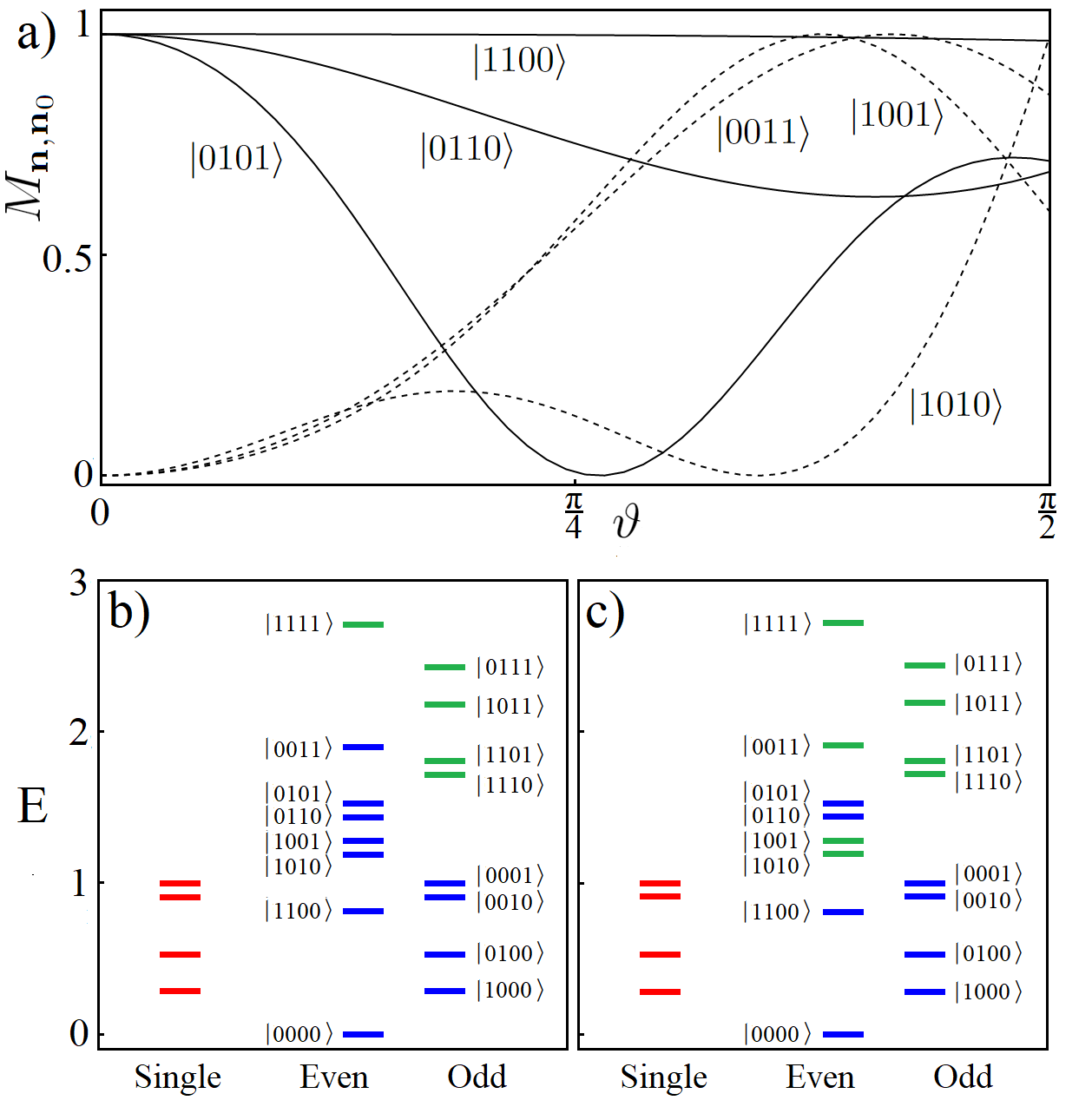}
    \caption{\new{Transition rates as well as single- and many-body Andreev states for a Josephson junction with the parameters in Fig.~\ref{fig6} and finite SOI and Zeeman field.
    (a) Transition rates $M_{{\bf n},{\bf n}_0}$ from the ground state $|{\bf n}_0\rangle=|0000\rangle$ to each of the six two-quasiparticle states (cf.~the first row of Table \ref{matrix}) vs the angle $\vartheta$ between the SOI axis and the Zeeman field. We only show non-zero transition rates within the same parity sector, where each rate is normalized to its maximum value in order to allow for a comparison of their $\vartheta$-dependence. The target state $|{\bf n}\rangle$ is specified near each curve. Solid (dashed) lines correspond to transitions which are allowed (forbidden) by selection rules for $\vartheta\to 0$. (b) Single and many-body states for ${\bf b}$ in the $x$-direction, i.e., for $\vartheta=\pi/2$, cf. Fig.~\ref{fig6}(c), with phase difference $\varphi_0=1.08\pi$. The red levels show the positive-energy single-particle ABS levels. The many-body states with zero (non-zero) transition rate $M_{{\bf n},{\bf n}_0}$ from the ground state are shown as green (blue) levels. (c) Same as  panel (b) but with the Zeeman field along the $z$-direction, i.e., for $\vartheta=0$, cf.~Fig.~\ref{fig6}(d).}
    }
    \label{fig7}
\end{center}
\end{figure}

\new{In Fig.~\ref{fig7}(a), we plot the transition rates $M_{{\bf n},{\bf n}_0}$ from the ground state $|{\bf n}_0\rangle$ to the six two-quasiparticle states ($n_1+n_2+n_3+n_4=2$) as a function of the angle $\vartheta$ between the Zeeman field and the polar SOI axis (for simplicity, $b_y=0$). We observe that, while the transition rates to the states $\ket{0110}$, $\ket{0101}$, and $\ket{1100}$ are different from zero for all $\vartheta$, the transition rates to the states $\ket{1010}$, $\ket{1001}$, and $\ket{0011}$ vanish for $\vartheta\to 0$ (i.e., for ${\bf b}$ along the $z$-direction). This is a signature for the onset of a selection rule.}

\new{After applying the pulse, the respective initial ($t=0$) population probabilities are then given by} $P_{{\bf n}_0}(0)=P_{{\bf n};{\rm stat}}$ and 
$P_{{\bf n}}(0)=P_{{\bf n}_0;{\rm stat}}$, while for all other states we have $P_{{\bf n}'}(0)=P_{{\bf n'};{\rm stat}}$.
We then solve Eq.~\eqref{matrixrate} subject to this initial condition.  For $\gamma_{\rm SO}^{}=0$ and ${\bf b}=0$, and using the transfer matrix in Ref.~\cite{Ackermann2023}, our scheme precisely
reproduces the results of Ref.~\cite{Ackermann2023} on dynamical parity stabilization after a microwave pulse. 
Below we study how the interplay of SOI and Zeeman field influences this phenomenon. For clarity, we focus on
the system parameters corresponding to panels (c) and (d) in Fig.~\ref{fig6}.

\begin{figure}
\begin{center}
    \includegraphics[width=0.48\textwidth]{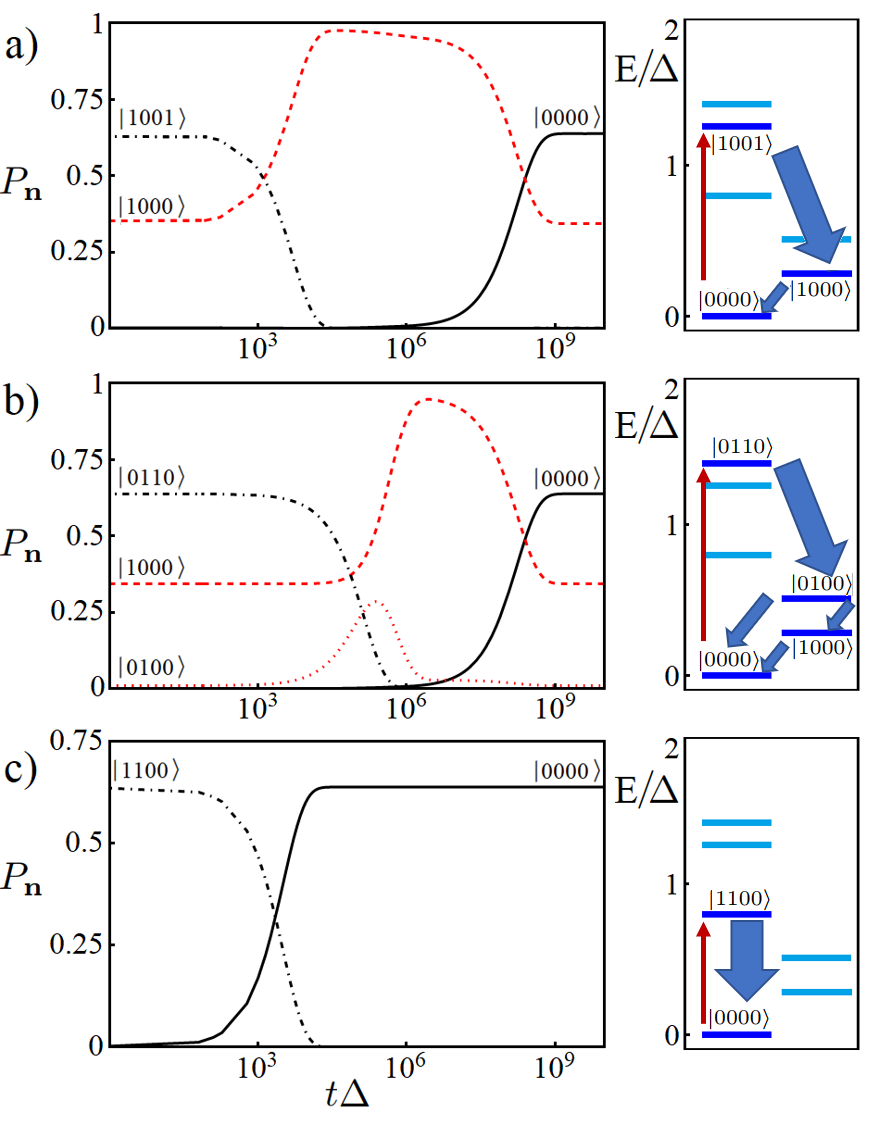}
    \caption{Many-body Andreev state population dynamics $P_{\bf n}(t)$ vs time (in units of $\Delta^{-1}$) for a Josephson junction with the parameters in Fig.~\ref{fig6}(c), with ${\bf b}$ in the
    $x$-direction and phase difference $\varphi_0=1.08\pi$.  We use $\kappa_0 = 0.1,\, \Omega = 10^{-3}\Delta,$ and $\eta_d = 0.01 \Delta$ in the
    spectral density \eqref{specdens}, with the background Ohmic part determined by $\alpha_0 = 10^{-4}$ and $\omega_c = \Delta$. We use the temperature scales
    $T_{\text{qp}} = 0.15\Delta$ and  $T_{\text{env}} = 0.07\Delta$. 
    Note the logarithmic time scales. We show only the curves for many-body levels with time-dependent probability weights. (There is some time-independent probability weight in other levels.)  Red curves correspond to odd-parity states, black curves to even-parity states.
    Three different transitions are shown in panels (a), (b), and (c), respectively, 
    which can be induced by a microwave drive from the ground state $|0000\rangle$.
    These transitions are shown as red arrows in the corresponding right column panels, where selected many-body Andreev energy levels are depicted; we again distinguish even- and odd-parity states, cf.~the right panel of Fig.~\ref{fig3}.  
    Thick (thin) arrows indicate large (small) transition rates connecting many-body Andreev states, cf.~Table \ref{matrix}. 
    }
    \label{fig8}
\end{center}
\end{figure}

We start with the case shown in Fig.~\ref{fig6}(c), where the Zeeman field is oriented along the 
nanowire direction. We note in passing that this configuration is typically considered for the generation of Majorana bound 
states at the nanowire ends \cite{Alicea2012}.   Applying a resonant microwave drive, one can then drive six different 
transitions out of the ground state $|{\bf n}_0\rangle$. \new{In Fig.~\ref{fig7}(b), the corresponding single-particle and many-body Andreev states are shown. Starting from the ground state, all  transitions to states with one or two quasiparticles have a non-zero transition rate  $M_{{\bf n},{\bf n}_0}$. However, 
only transitions to states with the same parity can be induced by the microwave pulse.} In Fig.~\ref{fig8}, we show the population dynamics after three of these microwave-induced transitions. In \new{Fig.~\ref{fig8}(a)}, we consider a resonant transition $|0000\rangle\to |1001\rangle$. However, 
driving the transition $|0000\rangle\to |1010\rangle$ instead gives very similar results.  
In \new{Fig.~\ref{fig8}(b),} we consider the resonant microwave-induced transition 
$|0000\rangle\to |0110\rangle$, where we obtain similar results for the population dynamics after the transition $|0000\rangle\to  |0101 \rangle$.
Finally, in \new{Fig.~\ref{fig8}(c),} we show the population dynamics after the transition $|0000\rangle\to |1100\rangle$, where
the transition $|0000\rangle\to |0011\rangle$ gives similar results.  We observe that for the cases shown in \new{Fig.~\ref{fig8}(a,b)}, 
an \emph{odd-parity} state (either $|0100\rangle$ or $|1000\rangle$) is occupied with large probability for a long intermediate time interval.
These observations correspond to the dynamical parity stabilization discovered in Ref.~\cite{Wesdorp2023}:
\emph{By driving a transition in the even-parity sector, one stabilizes the odd-parity polarization.} 
\new{It is worth noting that in Fig.~\ref{fig8}(b), there is a transition between both odd-parity states, with state $|0100\rangle$ acting as an intermediate state towards $|1000\rangle$. This behavior is a consequence of the level splitting induced by both SOI and Zeeman field. Such effects can play a crucial role in further increasing the lifetime of the odd-parity polarization effect. Indeed, the energy difference between the states $|0110\rangle$ and $|0100\rangle$ (which belong to different parity sectors)
is much bigger compared to the one between $|0100\rangle$ and $|1000\rangle$ (within the same parity sector). By suitably designing the electromagnetic environment such that the spectral density $J(\omega)$ exhibits a sub- or super-Ohmic behavior \cite{Weiss_2012} could allow one to modify the ratio $M_{|0100\rangle,|0110\rangle}/M_{|1000\rangle,|0100\rangle}$.  In that way, one may be able to further stabilize the lifetime of transient states as discussed, for example, in Ref.~\cite{Nava_2022}.}
We note that it is also possible to drive the system by a microwave pulse connecting two states in the odd-parity sector, and to thereby polarize the even-parity sector, but we do not discuss this case here.

In our model, the reason for the dynamical stabilization is the existence of a large many-body transition rate
into the respective odd-parity many-body state, cf.~Table~\ref{matrix}.  The largeness of the rate can be understood
from the closeness of some ABSs to the quasiparticle continuum. 
At the same time, the transition rate from the odd-parity state into the even-parity ground state $|0000\rangle$ is very small since 
all relevant ABSs are far away from the quasiparticle continuum.  This mechanism can explain the stabilization of the
odd-parity polarization at intermediate time scales \cite{Ackermann2023}. 
However, for the transition in Fig.~\ref{fig8}(c), the vanishing rate from the excited even-parity state into the intermediate 
odd-parity state excludes this phenomenon.  We conclude from Fig.~\ref{fig8} that the combined effects of SOI and Zeeman
field may result in qualitative changes in the many-body population dynamics in the Andreev sector.  \new{Indeed, the energy splitting induced by the SOI and the Zeeman field allows one to have a non-zero spectral density \eqref{specdens}, and thus a non-zero transition rate between states which are otherwise disconnected. At the same time, selection rules are less restrictive 
due to the fact that orbital and spin angular momenta are no longer conserved. As a consequence, a wider set of initial conditions
can be explored, exhibiting different parity polarization behavior depending on precisely which transition is driven.}

\begin{figure}
\begin{center}
    \includegraphics[width=0.48\textwidth]{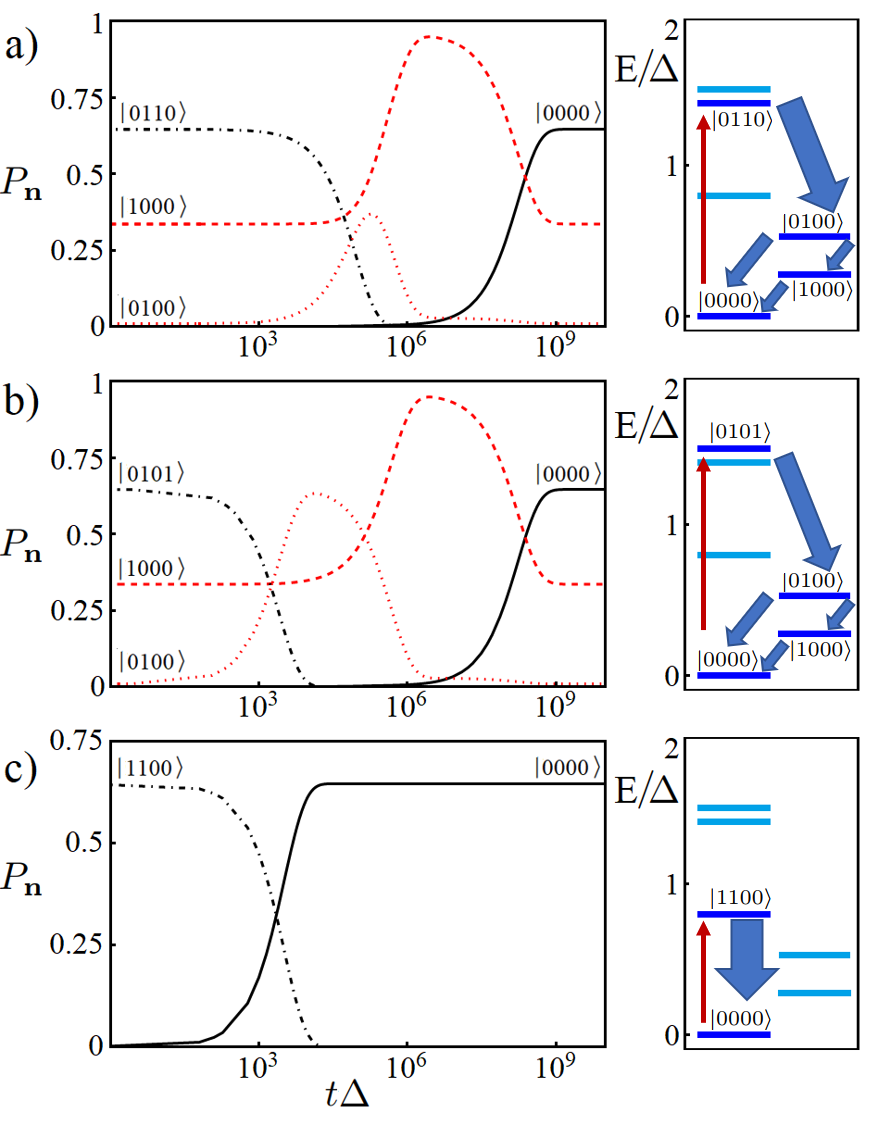}
    \caption{Many-body Andreev state population dynamics $P_{\bf n}(t)$ vs time (in units of $\Delta^{-1}$) for the parameters in Fig.~\ref{fig6}(d), with ${\bf b}$ in the $z$-direction
    and the phase difference $\varphi_0=1.08\pi$. The spectral density of the environment was taken as in Fig.~\ref{fig8}.
    Note the logarithmic time scales. We show only the curves for many-body levels with time-dependent probability weights. (There is some time-independent probability weight in other levels.)
    Red curves correspond to odd-parity states, black curves to even-parity states.
    In contrast to the case in Fig.~\ref{fig6}, here only three transitions, corresponding to panels (a), (b), and (c), can be induced by a microwave drive starting from the ground state $|{\bf n}_0\rangle$ because of selection rules. These transitions are shown as red arrows in the corresponding right column panels, where selected many-body Andreev energy levels are depicted.  
    Thick (thin) arrows indicate large (small) many-body transition rates.    }
    \label{fig9}
\end{center}
\end{figure}

Next we turn to the parameter choice corresponding to Fig.~\ref{fig6}(d), where the Zeeman field is oriented along the $z$-direction.
We then obtain the population dynamics shown in the left column of Fig.~\ref{fig9}, where the three panels (a), (b), and (c) correspond
to the three  possible transitions from the ground state $|{\bf n}_0\rangle$ which can be induced by a resonant microwave field and 
which are allowed by selection rules.  
\new{In Fig.~\ref{fig7}(c), we show the single-particle and many-body Andreev states for $\vartheta=0$, where the Zeeman field is aligned along the $z$-direction. As in Fig.~\ref{fig7}(b), we have highlighted all states that exhibit a non-zero transition rate $M_{{\bf n},{\bf n}_0}$ with the ground state.}
This is in contrast to the case shown in Fig.~\ref{fig8} with a Zeeman field in the
$x$-direction \new{($\vartheta=\pi/2$)}, where six transitions are allowed by selection rules but we show only three of those. 
For the microwave-induced transitions shown in panels (a) and (b) of Fig.~\ref{fig9}, we again observe a dynamical polarization
of the odd-parity sector at intermediate times, where two odd-parity states are relevant.
For the transition shown in panel (c), we once more encounter a case where a vanishing transition rate into the odd-parity state excludes dynamical parity polarization.    The qualitative impact of SOI and Zeeman field on this phenomenon is therefore of similar importance as for the case shown in Fig.~\ref{fig8}.

\begin{figure}
\begin{center}
    \includegraphics[width=0.48\textwidth]{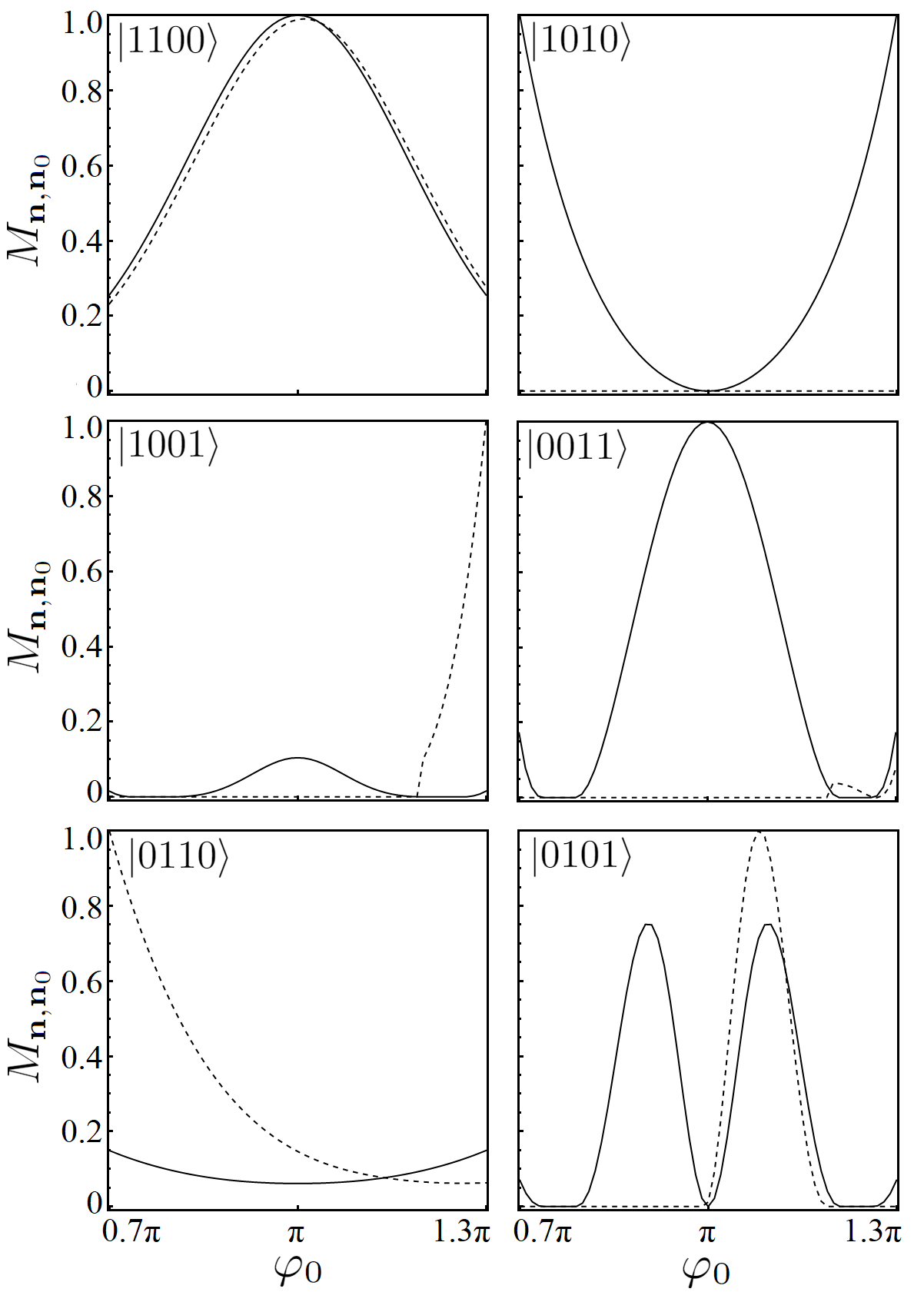}
    \caption{\new{Transition rates $M_{{\bf n},{\bf n}_0}$ from the ground state $|{\bf n}_0\rangle$ to each of the six possible two-quasiparticle states $|{\bf n}\rangle$ vs phase difference $\varphi_0$ for a Josephson junction with the parameters in Fig.~\ref{fig6}. The target states $|{\bf n}\rangle$ are shown in each panel. Solid (dashed) lines correspond to the angle $\vartheta=\pi/2$ ($\vartheta=0$) between ${\bf b}$ and the SOI polar axis. In each panel, transition rates are normalized to  their maximum value in the shown interval.} 
    }
    \label{fig10}
\end{center}
\end{figure} 

\new{In the absence of the SOI and the Zeeman field,  selection rules can be inferred by analyzing the orbital and spin angular momenta of each ABS. When spin degeneracy is broken, however, the transition rates exhibit a non-trivial dependence on both $\vartheta$, see Fig.~\ref{fig7}(a), and on the phase difference $\varphi_0$. In Fig.~\ref{fig10}, we show the $\varphi_0$-dependence of the transition rates $M_{{\bf n},{\bf n}_0}$ from the ground state $|{\bf n}_0\rangle$ to each of the six possible two-quasiparticle states $|{\bf n}\rangle$, both for $\vartheta=0$ and for $\vartheta=\pi/2$. Similarly to panels (c) and (d) of Fig.~\ref{fig6}, the transition rates for $\vartheta=\pi/2$ are symmetric around $\varphi_0=\pi$, while for $\vartheta=0$, a strong asymmetry is present. Furthermore, the transition rates quickly drop to zero for some values of the phase difference, 
pointing out the onset of a selection rule for the corresponding target states. Compared to the case without SOI and Zeeman field, by properly tuning $\varphi_0$, one can thus select which states can be accessed by an external perturbation. 
In agreement with Figs.~\ref{fig8} and \ref{fig9}, for $\varphi_0=1.08\pi$ and $\vartheta=\pi/2$, all six transitions rates are different from zero, while three of them vanish for $\vartheta\to 0$.}

\begin{figure}
\begin{center}
    \includegraphics[width=0.48\textwidth]{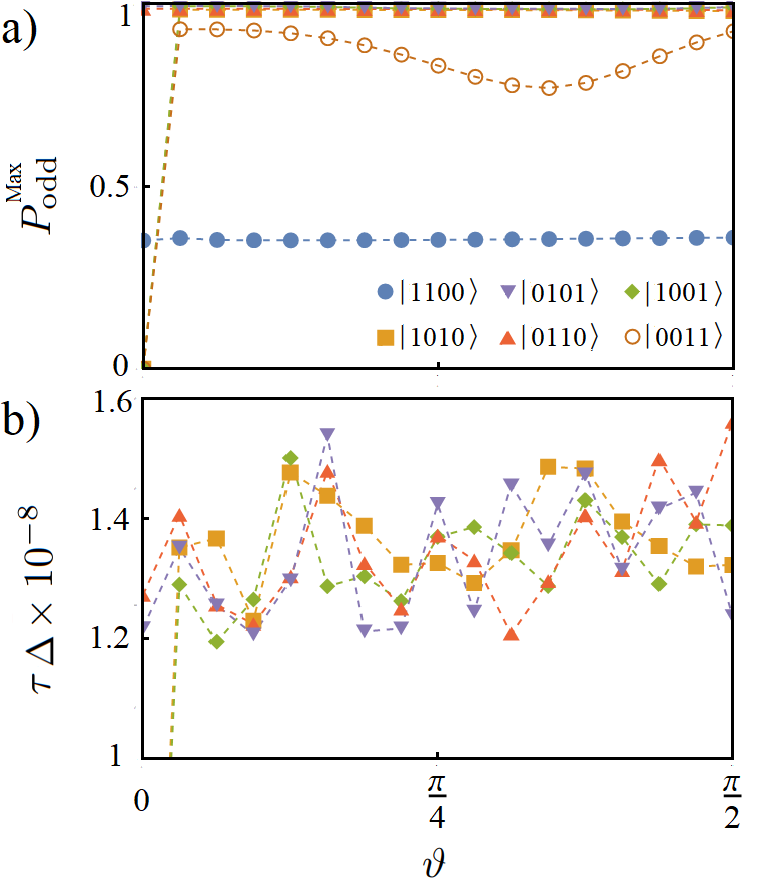}
    \caption{Dynamical parity polarization as a function of the angle $\vartheta$ between the Zeeman field and the polar SOI axis.
    The spectral density of the environment was taken as in Fig.~\ref{fig8}. (a)
    Maximally achievable odd-parity polarization $P^{\rm Max}_{\rm odd}$ vs $\vartheta$ after each of the six possible microwave-induced
    transitions $|{\bf n_0}\rangle\to |{\bf n}\rangle$. We use the system parameters corresponding to Fig.~\ref{fig8} (where $\vartheta=\pi/2$) and Fig.~\ref{fig9} (where $\vartheta=0$).
    Dashed lines are guides to the eye only.  Different symbols correspond to the excited initial states $|{\bf n}\rangle$, as explained in the legend.
    (b) Odd-parity lifetime $\tau$ (in units of $10^{8}\Delta^{-1}$) vs $\vartheta$ for the six transitions in panel (a).  For two of these transitions, the lifetime is very short and does not appear on the scale of the figure.  Dashed lines are guides to the eye only. The symbols are used as in panel (a).  
    }
    \label{fig11}
\end{center}
\end{figure} 

In Fig.~\ref{fig11}, we study how the dynamical parity polarization effect depends on the angle $\vartheta$.  We examine all six transitions $|{\bf n}_0\rangle\to |{\bf n}\rangle$ that can in principle be excited by a resonant microwave driving pulse. For each $|{\bf n}\rangle$, we  
determine the maximal probability for occupying the odd-parity sector during the time evolution, $P_{\rm odd}^{\rm Max}$, and the lifetime of the corresponding odd-parity
states, $\tau$.  We define the latter time scale as the half-width of the corresponding broad peak in $P_{\rm odd}(t)$, see Figs.~\ref{fig8} and \ref{fig9}.
We observe from Fig.~\ref{fig11}(a) that the achievable odd-parity polarization depends significantly on which transition is driven, while there is only a weak dependence 
on the angle $\vartheta$ (except near $\vartheta=0$).    
Importantly, almost full odd-parity polarization is possible for several resonant drive frequencies while for other drive frequencies, the system does not
get polarized at all, see Fig.~\ref{fig8}(c) and Fig.~\ref{fig9}(c).  As shown in Fig.~\ref{fig11}(b), the lifetime $\tau$ of the odd-parity polarization state is rather insensitive 
of the angle $\vartheta$ as long as one chooses one of the drive frequencies corresponding to large $P_{\rm odd}^{\rm Max}$.  The  variations in Fig.~\ref{fig11}(b) come from changes in the transition rates $\Gamma_{\mu,\nu}$ with $\vartheta$.  We conclude that the combined effect of SOI and Zeeman fields can influence the dynamical polarization effect, 
both concerning the degree of polarization and (to a lesser degree) the achievable lifetimes. 

\section{Conclusions}\label{sec5}

In this work, we have put forward a theoretical approach for describing the many-body quantum dynamics of superconducting systems with spin-orbit coupling and magnetic fields. 
It is well known   that such systems can be efficiently described in terms of a doubled
Nambu spinor approach, where one keeps  the electron and hole spinors with both spin projections.   This doubling of the actual number of degrees of freedom is referred to as double-counting problem and can give rise to spurious many-body effects if the theory is constructed in a cavalier manner.   We resolve this general problem by working in the so-called excitation picture, where only the positive single-particle solutions of the BdG equation are employed to construct the many-body theory.  This is possible since the negative-energy solutions are related to the corresponding 
positive-energy solutions by particle-hole symmetry, and we systematically exploit this relation in our approach.  

We apply our general formalism to a Josephson junction formed by a clean 1D nanowire with spin-orbit coupling in a Zeeman field, 
which is tunnel-coupled at its end to superconducting banks.  The junction is embedded in a loop and inductively coupled to a microwave resonator, see Fig.~\ref{fig1}.  In the absence of the electromagnetic environment defined by the resonator, the BdG single-particle problem can be solved exactly.  This solution provides a 
convenient basis for the construction of many-body states.  From the diagonal elements of the reduced density operator of the many-body 
Andreev bound states, the Lindblad equation derived in Sec.~\ref{sec3} then yields a matrix rate equation for the population dynamics 
of the corresponding many-body Andreev  states.  We here study how the corresponding populations evolve in time after a strong initial microwave pulse driving a specific transition.  This question is related to the dynamical parity stabilization phenomenon discovered experimentally in Ref.~\cite{Wesdorp2023}.  Previous results \cite{Ackermann2023}  for the simpler case without spin-orbit coupling and without Zeeman field 
are recovered by our results.  We find that, depending on the microwave driving frequency, the maximally reachable parity polarization $P_{\rm odd}^{\rm Max}$ and, to a lesser degree, the 
time scale over which the odd-parity sector becomes dynamically stabilized, show a dependence on the angle $\vartheta$ between the spin-orbit polar axis and the Zeeman field.   Our results suggest
that one can optimize the parity stabilization mechanism by proper field alignment.  

To conclude, we have introduced a systematic theoretical framework for studying the quantum many-body dynamics of superconducting systems where a doubling of the fermionic space is indicated, e.g., due to the presence of spin-orbit interactions and 
Zeeman fields.  The presence of particle-hole symmetry then implies that the excitation picture allows for the construction of
a many-body theory free from the double-counting problem.  
We believe that the approach proposed here will be useful also for many other theoretical many-body studies in the future.


\begin{acknowledgments}
We thank Nico Ackermann, Domenico Giuliano, and Alfredo Levy Yeyati for discussions. We acknowledge funding by the Deutsche Forschungsgemeinschaft (DFG, German Research Foundation) Grant No.~277101999 - TRR 183 (project C01) and under Germany's Excellence Strategy - Cluster of Excellence Matter and Light for Quantum Computing (ML4Q) EXC 2004/1 - 390534769.  
\end{acknowledgments}


\appendix

\section*{Appendix: BdG solutions}  

In this Appendix, we summarize the solution of the BdG problem defined by Eq.~\eqref{bdg} and the 
supercurrent matrix elements in Eq.~\eqref{JC}.  For simplicity,
we set $\Delta=v_F = 1$ below.  

\emph{Andreev bound states.---}We begin with ABS solutions ($\mu=\lambda$) with dispersion $E=E_\lambda(\varphi_0)$. 
For $|E|<\Delta$, with the Heaviside step function $\Theta(x)$ and the Nambu spinor form in Eq.~\eqref{Nambu}, ABS solutions of the BdG equation are given by
\begin{eqnarray} \label{ABSwf}
    \Phi_{E, \lambda}(x) &=& \frac{\Theta(-x) e^{\kappa_\lambda x}}{\sqrt{2}} 
    \begin{pmatrix}
        a_{\lambda} e^{-i\gamma_{\lambda}/2} \\
        b_{\lambda} e^{i\gamma_{\lambda}/2} \\
        a_{\lambda} e^{i\gamma_{\lambda}/2} \\
        b_{\lambda} e^{-i\gamma_{\lambda}/2}   
    \end{pmatrix} \\ \nonumber
    &+& \frac{\Theta(x) e^{-\kappa_\lambda x}}{\sqrt{2}} 
    \begin{pmatrix}
        c_{\lambda} e^{i\gamma_{\lambda}/2} \\
        d_{\lambda} e^{-i\gamma_{\lambda}/2} \\
        c_{\lambda} e^{-i\gamma_{\lambda}/2} \\
        d_{\lambda} e^{i\gamma_{\lambda}/2} 
    \end{pmatrix},
\end{eqnarray}
where $E_{\lambda} = \cos\gamma_{\lambda}$ and $\kappa_{\lambda}=\sin\gamma_{\lambda}$. We choose
 $\gamma_{\lambda} \in (0, \pi)$ and use $a_\lambda= (a_{\lambda,\uparrow},a_{\lambda,\downarrow})^T$, and similarly for $b_\lambda,c_\lambda,$ and $d_\lambda$. 
 The normalization condition for the amplitudes in Eq.~\eqref{ABSwf} is 
\begin{equation}
    \sum_{\sigma} \left (|a_{\lambda, \sigma}|^2 + |b_{\lambda, \sigma}|^2 +|c_{\lambda, \sigma}|^2 + |d_{\lambda, \sigma}|^2 \right) = \frac{2\kappa_{\lambda}}{1+\kappa_{\lambda} L}.
\end{equation}
The ABS dispersion relation follows by inserting Eq.~\eqref{ABSwf} into the matching condition
\eqref{matchingfinal}.  
Nontrivial solutions require the vanishing of a corresponding determinant, which leads to Eq.~\eqref{detzero}. 
The corresponding eigenvectors then determine the ABS wave functions.

\emph{Continuum states.---}Quasiparticle continuum states with energy $|E|>\Delta$ are labeled by the multi-index $p = (E, s, \sigma)$, with the scattering channel index $s \in \{1,2,3,4\}$ \cite{Ackermann2023} and the spin index $\sigma$. The corresponding Nambu states are given by a sum of an incoming and a scattered outgoing state,
$\Phi_{p}(x) = \Phi_{p}^{(\text{in})}(x) + \Phi_{p}^{(\text{out})}(x).$
With $\sigma_E={\rm sgn}(E)$ and the length ${\cal L}$ of the superconducting bank, an incoming state of type $s$ can be written as 
\begin{widetext}
\begin{equation}
\Phi_{p}^{(\text{in})}(x)=\frac{\Theta(-x)}{\sqrt{2\cosh \tilde \gamma}} \frac{e^{ikx}}{\sqrt{\mathcal{L}}}  
\begin{pmatrix} \delta_{s,1}\delta_{\sigma,\uparrow} \, e^{\tilde\gamma/2} \\
  \delta_{s,1}\delta_{\sigma,\downarrow}\,
    e^{\tilde\gamma/2} \\  \delta_{s,2}\delta_{\sigma,\uparrow}
      \,   e^{-\tilde\gamma/2}  \\ \delta_{s,2}\delta_{\sigma,\downarrow}
    \, e^{-\tilde\gamma/2} \\ \delta_{s,1}\delta_{\sigma,\uparrow}
      \, \sigma_{E} e^{-\tilde\gamma/2} \\ \delta_{s,1}\delta_{\sigma,\downarrow}
\, \sigma_{E}  e^{-\tilde\gamma/2} \\ \delta_{s,2}\delta_{\sigma,\uparrow}
         \, \sigma_E  e^{\tilde\gamma/2} \\ \delta_{s,2}\delta_{\sigma,\downarrow}
         \, \sigma_E e^{\tilde\gamma/2}
    \end{pmatrix}+ \frac{\Theta(x)}{\sqrt{2\cosh \tilde \gamma}}\frac{e^{-ikx}}{\sqrt{\mathcal{L}}} 
\begin{pmatrix} \delta_{s,3}\delta_{\sigma,\uparrow}\, e^{-\tilde\gamma/2} \\
   \delta_{s,3}\delta_{\sigma,\downarrow}  \, e^{-\tilde\gamma/2} \\
    \delta_{s,4}\delta_{\sigma,\uparrow}   \,  e^{\tilde\gamma/2}  \\
   \delta_{s,4}\delta_{\sigma,\downarrow}\, e^{\tilde\gamma/2} \\
   \delta_{s,3}\delta_{\sigma,\uparrow}\, \sigma_{E} e^{\tilde\gamma/2} \\
     \delta_{s,3}\delta_{\sigma,\downarrow}  \, \sigma_{E}  e^{\tilde\gamma/2} \\
     \delta_{s,4}\delta_{\sigma,\uparrow}    \, \sigma_E  e^{-\tilde\gamma/2} \\
      \delta_{s,4}\delta_{\sigma,\downarrow}  \, \sigma_E e^{-\tilde\gamma/2}
    \end{pmatrix},
\end{equation}
where $|E|=\cosh\tilde\gamma$ with $\tilde\gamma(E)\in [0,\infty)$ and  $k(E)=\sigma_E\sinh\tilde\gamma(E)$.
Similarly, for a given incident (incoming) state with quantum numbers $p$,
the scattered (outgoing) state is written as
\begin{equation}
  \Phi_{p}^{(\text{out})}(x) =\frac{\Theta(-x)}{\sqrt{2\cosh \tilde \gamma}}  \frac{e^{ikx}}{\sqrt{\mathcal{L}}} 
    \begin{pmatrix}
        a_{p, \uparrow}\,e^{-\tilde \gamma/2} \\
        a_{p, \downarrow}\,e^{-\tilde \gamma/2} \\
        b_{p, \uparrow}\,e^{\tilde \gamma/2} \\
        b_{p, \downarrow}\,e^{\tilde \gamma/2}\\
        a_{p, \uparrow}\, \sigma_E e^{\tilde \gamma/2} \\
        a_{p, \downarrow}\,\sigma_E e^{\tilde \gamma/2} \\
        b_{p, \uparrow}\,\sigma_E e^{-\tilde \gamma/2} \\
        b_{p,  \downarrow}\,\sigma_E e^{-\tilde \gamma/2}  
    \end{pmatrix}+   \frac{\Theta(x)}{\sqrt{2\cosh \tilde \gamma}}  \frac{e^{-ikx}}{\sqrt{\mathcal{L}}} 
 \begin{pmatrix}
        c_{p,\uparrow}\,e^{\tilde \gamma/2} \\
        c_{p,  \downarrow}\,e^{\tilde \gamma/2} \\
        d_{p,  \uparrow}\,e^{-\tilde \gamma/2} \\
        d_{p,  \downarrow}\,e^{-\tilde \gamma/2}\\
        c_{p, \uparrow}\,\sigma_E e^{-\tilde \gamma/2} \\
        c_{p, \downarrow}\,\sigma_E e^{-\tilde \gamma/2} \\
        d_{p, \uparrow}\,\sigma_E e^{\tilde \gamma/2} \\
        d_{p, \downarrow}\,\sigma_E e^{\tilde \gamma/2} 
    \end{pmatrix}.
\end{equation}
The normalization condition for the complex-valued 
scattering amplitudes ($a_{p,\sigma},b_{p,\sigma},c_{p,\sigma},d_{p,\sigma})$ is given by
\begin{equation}
\sum_{\sigma}\left(|a_{p, \sigma}|^2+|b_{p, \sigma}|^2 + |c_{p,  \sigma}|^2 + |d_{p, \sigma}|^2\right) = 1.
\end{equation}
One can then determine the scattering amplitudes, and thereby the quasiparticle wave functions, by inserting the 
above Ansatz into the matching condition \eqref{matchingfinal}.  
This implies a linear algebra problem that can easily be solved numerically.

\emph{Current matrix elements.---}Next we discuss the matrix elements $\mathcal{I}_{\mu,\nu}$ in Eq.~\eqref{JC}.
First, if both indices $(\mu,\nu)=(\lambda,\lambda')$ correspond to ABSs, we obtain 
\begin{equation}
    \mathcal{I}_{\lambda,\lambda'} =\frac{\frac{E_{\lambda}-E_{\lambda'}}{2}\sin\left(\frac{\gamma_{\lambda}-\gamma_{\lambda'}}{2}\right) + \sin\left(\frac{\gamma_{\lambda}+\gamma_{\lambda'}}{2} \right)}{\kappa_{\lambda}+\kappa_{\lambda'}}\times
    \sum_{\sigma}\left(a^*_{\lambda, \sigma}a^{}_{\lambda', \sigma} - b^*_{\lambda, \sigma}b^{}_{\lambda', \sigma} + c^*_{\lambda, \sigma}c_{\lambda', \sigma}^{} - d^*_{\lambda, \sigma}d^{}_{\lambda', \sigma} \right).
\end{equation}
Second, following similar arguments as in Ref.~\cite{Ackermann2023}, we find that for ${\cal L}\to \infty$, all current matrix elements between continuum states $(\mu,\nu)=(p,p')$ vanish, $\mathcal{I}_{p,p'}=0$. 
Superconducting phase fluctuations hence do not induce transitions between continuum states. 
Finally, for transitions between an ABS with energy $E_{\lambda}$ and a continuum state with quantum numbers $p=(E,s,\sigma)$, we obtain
\begin{eqnarray}\nonumber
    \mathcal{I}_{\lambda,p} &=& \frac{i}{\sqrt{\mathcal{L} \cosh \tilde \gamma}}  \sum_{\sigma'}
    \biggl[\frac{1}{\kappa_{\lambda}-ik} \left\{\left(a^*_{\lambda, \sigma'} a^{}_{p,\sigma'} - 
    d^*_{\lambda, \sigma'}d^{}_{p,\sigma'}\right) W(-z)
    +\left(b^*_{\lambda, \sigma'}b^{}_{p,\sigma'} - c^*_{\lambda, \sigma'}c_{p,\sigma'}^{}\right)W(z)\right\} 
     \\ & +&\frac{\delta_{\sigma',\sigma}}{\kappa_{\lambda}+ik}\left\{\left(a^*_{\lambda, \sigma} \delta_{s,1}- 
    d^*_{\lambda, \sigma} \delta_{s,4} \right)W(z^*) 
    \left(b^*_{\lambda, \sigma} \delta_{s,2}- c^*_{\lambda, \sigma} \delta_{s,3}\right)W(-z^*)
    \right\}
    \biggr],
\end{eqnarray}
\end{widetext}
where we use 
\begin{eqnarray} \nonumber
    W(z) &=& w(z) + \frac{E-E_{\lambda}}{2}w^*(z),\\
    w(z) &=& \Theta(E)\sinh(z)+\Theta(-E)\cosh(z),\\ \nonumber
    z &=& (\tilde\gamma(E)+i\gamma_\lambda)/2.
\end{eqnarray}
We note that for $\mathcal{L} \rightarrow \infty$, summations over $p = (E, s, \sigma)$ can be performed by using
\begin{equation}
    \frac{1}{\mathcal{L}} \sum_{p} (\cdots) = \int dE\, \nu(E) \sum_{s=1}^4 \sum_{\sigma} (\cdots),
\end{equation}
where $\nu(E)$ is the BCS density of states (per unit length, and recalling our convention $\Delta=v_F=1$),
\begin{equation}
    \nu(E) = \frac{1}{2\pi}\frac{|E|}{\sqrt{E^2-1}} \Theta(|E|-1).
\end{equation}

\bibliography{biblio}
\end{document}